\documentclass[12pt]{iopart}
\usepackage{epsf}

\begin{document}
\newcommand{\bd     }{\begin{displaymath}}
\newcommand{\ed     }{\end{displaymath}}
\newcommand{\plus   }{{\!+\!}}
\newcommand{\minus  }{{\!-\!}}
\newcommand{\order  }{{\cal O}}
\newcommand{\vsp    }{\vspace*{3mm}}
\newcommand{\s      }{\sigma}
\newcommand{\be     }{\beta}
\newcommand{\ra     }{\rightarrow}
\newcommand{\ua     }{\uparrow}
\newcommand{\da     }{\downarrow}
\newcommand{\mc     }{\mathcal}
\newcommand{\bra    }{\langle}
\newcommand{\ket    }{\rangle}
\newcommand{\Bra    }{\left\langle}
\newcommand{\Ket    }{\right\rangle}
\newcommand{\mb     }{\mbox{\boldmath$m$}}
\newcommand{\mh     }{\hat{m}}
\newcommand{\la     }{\lambda}
\newcommand{\sech   }{\mathrm{sech}}
\newcommand{\bM     }{\mbox{\boldmath$M$}}
\newcommand{\bA     }{\mbox{\boldmath$A$}}
\newcommand{\bB     }{\mbox{\boldmath$B$}}
\newcommand{\bF     }{\mbox{\boldmath$F$}}
\newcommand{\bP     }{\mbox{\boldmath$P$}}
\newcommand{\bv     }{\mbox{\boldmath$V$}}
\newcommand{\bh     }{\mbox{\boldmath$h$}}
\newcommand{\bw     }{\mbox{\boldmath$w$}}
\newcommand{\bsigma }{\mbox{\boldmath$\sigma$}}
\newcommand{\bXi    }{\mbox{\boldmath$\Xi$}}
\newcommand{\bT     }{\mbox{\boldmath$T$}}
\newcommand{\bG     }{\mbox{\boldmath$G$}}
\newcommand{\bD     }{\mbox{\boldmath$D$}}
\newcommand{\bSigma }{\mbox{\boldmath$\Sigma$}}
\newcommand{\bsi    }{\mbox{\boldmath$\sigma$}}
\newcommand{\bolds  }{\mbox{\boldmath$s$}}
\newcommand{\bt     }{\mbox{\boldmath$T$}}
\newcommand{\bdeltaL}{\mbox{\boldmath$\delta L$}}
\newcommand{\bone   }{\mbox{\boldmath$1$}}
\newcommand{\bomega }{\mbox{\boldmath$\omega$}}
\newcommand{\btau   }{\mbox{\boldmath$\tau$}}
\newcommand{\bk     }{\mbox{\boldmath$k$}}
\newcommand{\bpsi   }{\mbox{\boldmath$\psi$}}
\newcommand{\bz     }{\mbox{\boldmath$z$}}
\newcommand{\bZ     }{\mbox{\boldmath$Z$}}
\newcommand{\bzero  }{\mbox{\boldmath$0$}}
\newcommand{\bV     }{\mbox{\boldmath$V$}}
\newcommand{\bphi   }{\mbox{\boldmath$\phi$}}
\jl{1}

\title[Structure Formation in Random Hetero-Polymers]{A Solvable Model of Secondary Structure Formation in Random Hetero-Polymers}

\author{\bf N.S.~Skantzos, J.~van~Mourik and A.C.C.~Coolen}

\address{Department of Mathematics, King's College London\\
         The Strand, London WC2R~2LS, U.K.}

\date{\today}

\begin{abstract}
We propose and solve a simple model describing secondary structure formation in random hetero-polymers.
It describes monomers with a combination of one-dimensional short-range
interactions (representing steric forces and hydrogen bonds) and
infinite range interactions (representing polarity forces).
We solve our model using a combination of mean field and random field techniques, leading to
phase diagrams exhibiting second-order transitions between folded, partially folded and unfolded
states, including regions where folding depends on initial conditions.
Our theoretical results, which are in excellent agreement with numerical
simulations, lead to an appealing physical picture of the folding process:
the polarity forces drive the transition to a collapsed state, the steric
forces introduce monomer specificity, and the hydrogen bonds
stabilise the conformation by damping the frustration-induced
multiplicity of states.
\end{abstract}


\section{Introduction}

Proteins are polymeric chains of amino-acids.
The successful functioning of a protein in a living organism depends crucially, among
other factors, on its ability to fold into a desired three-dimensional structure
(its `native state'), and to subsequently attach in a very specific way to
other macro-molecules. From a biological and medical point of view, it is
therefore highly desirable to know which native state corresponds to a given
amino-acid sequence, and (conversely, for therapeutic purposes) to know which amino-acid
sequence would fold into a desired native state; this requires a quantitative understanding of
the physical forces underlying the folding
mechanism. A detailed identification of sequence-specific native states will
necessarily involve sophisticated (molecular dynamics based) computational methods.
However, due to the large number of degrees of freedom of proteins, the complicated nature of
the various types of electro-chemical interactions and the so-called `hard' geometric
chain constraints of a protein, such computer programmes are unfortunately (as yet) extremely slow.
Thus, in order to identify the role and degree of importance of the various folding
parameters,
 a theoretical (i.e. statistical mechanical) analysis would be very
 welcome.

It is generally assumed that the presently observed population of real proteins has
evolved from the larger class of random hetero-polymers, driven by natural selection.
This suggests that the study of random hetero-polymers is a
natural first step {\em en route} towards the
statistical mechanical study of proteins.
Furthermore, already at an early stage it was recognized \cite{bryngelson}, via a theoretical
study based on the random energy scheme \cite{derrida_REM}, that many aspects of
protein folding (such as the appearance of `mis-folded' phases, and transitions
between folded and unfolded states) can be understood on the basis of
equilibrium statistical mechanical calculations for random hetero-polymers.
Even simple models with only two types of amino-acids interacting with the water
solvent, viz. hydro-phobic amino-acids versus polar ones, can successfully describe the
basics of protein folding (see e.g. the so-called HP model \cite{lau_dill}). Further
statistical mechanical approaches include replica calculations on polymer chains with
Gaussian pair interactions \cite{garel_orland,shak_gutin}, variational analyses
in replica spaces \cite{variational1,variational2}, lattice models \cite{lattice1,lattice2} and
lattice gas models \cite{jort}, to mention but a few. In most of these examples, analytical solvability
relies on the absence of spatial structure, which allows for more or less
conventional mean-field statistical mechanics.

In this paper we extend the class of analytically solvable models in this field.
We present a model for secondary structure
formation in random hetero-polymers consisting of amino-acid monomers which are allowed to
interact in three qualitatively different ways: (i) via so-called steric interactions, which
reflect monomer-specific geometric constraints and electrical forces
determining the local energy landscape for the orientation of monomer-connecting
links, (ii) via hydrogen-bonding, which acts over larger distances along the chain,
and is believed to play a role in the stabilization of helix-type structures,
and (iii) via polarity-induced energy gradients, which tend to promote states in which
the hydrophobic amino-acids are more or less turned towards the same side of the
polymeric chain, in order to enable effective shielding from water molecules via folding
of the polymer as a whole. Interactions (i) and (ii) are of a short-range nature,
whereas (iii) is long-range.
We note that secondary
structure formation has also been studied
within a mean-field approach in \cite{archontis}, and that a combination of
different types of monomer interactions has been proposed previously in
\cite{bryngelson}. In the latter study, assuming statistical independence of
energy levels, the random energy scheme could provide qualitative results;
however, the validity of this approach has since then been questioned
\cite{pande}. In contrast, our solution does not employ random energy considerations.
It is based on a combination of mean-field and random transfer-matrix techniques, which
in  one-dimensional models are known to reduce the evaluation of the partition
function to a relatively simple numerical problem. Due to the presence
of additional long-range interactions (via polarity-induced forces) our model no longer
lies in the universality class of one dimensional systems, and phase transitions
are therefore possible (and will indeed occur) at finite temperatures.

Our paper is organized as follows.
We first define our model and the relevant
macroscopic observables. Since the disordered infinite-range (polarity induced) part of our Hamiltonian,
which drives the collapse to a folded state, is
different in structure from the more familiar Mattis-like \cite{mattis} terms in
mean-field spin systems,
we first solve our model for the case where only polarity energies are
present. We then proceed to the solution of the full
 model, with all three interaction terms present, but now limiting ourselves (for simplicity)  to the simplest choice
 of angular variables. Our phase diagrams exhibit
second-order transitions between folded and unfolded states, whereas close to
zero-temperature a hierarchy of `mixed' phases appears, where new ergodic
components are created and where folding depends on initial conditions. The latter phases
are found to be related to entropic discontinuities. Finally, we present results
from simulation experiments,  which show excellent agreement with the theory.

\section{Model Definitions}

\begin{figure}[t]
\setlength{\unitlength}{1mm}
\vsp
\begin{picture}(100,55)
\put( 50,  5){\epsfysize=50\unitlength\epsfbox{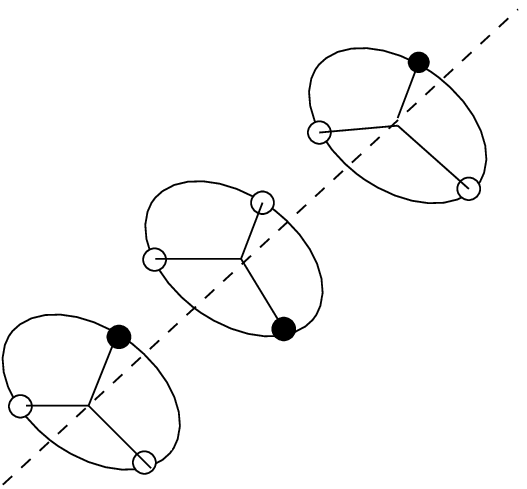}}
\put( 39, 12){$\phi=0$}
\put( 58, 25){$\frac{2\pi}{3}$}
\put( 61,  2){$-\frac{2\pi}{3}$}
\end{picture}
\vsp
\caption{Illustration of the physical meaning of our clock-state spin variables $\phi_i$.
A spin state $\phi$ represents the physical location of an individual monomer, relative to the
one-dimensional polymer chain axis (the `backbone', drawn as a dashed
line). In this graph the number of possible locations for any given monomer is $q=3$. The black blobs
represent locations occupied by a monomer. }
\label{fig:definition}
\end{figure}

We consider one dimensional models of random hetero-polymers, where $N$
clock-state spin variables  $\phi_i\in
\left\{\frac{(2k+1)\pi}{q}; k=0,\ldots,q-1\right\}$ describe the spatial
orientations of successive monomer residues in planes vertical to the polymer's chain axis, see
figure \ref{fig:definition}.
The configurational state of the system as a whole is written as
$\bphi=(\phi_1,\ldots,\phi_N)$.
%
We define the Hamiltionian of the system to be the sum  of three qualitatively different
terms,
$H(\bphi)=H_{\rm s}(\bphi)+H_{\rm p}(\bphi)+H_{\rm
Hb}(\bphi)$,
which are defined and interpreted as follows:
\begin{figure}[t]
\vspace*{10mm}
\setlength{\unitlength}{1.4mm}
\begin{picture}(100,55)
\put( 30,  0){\epsfysize=60\unitlength\epsfbox{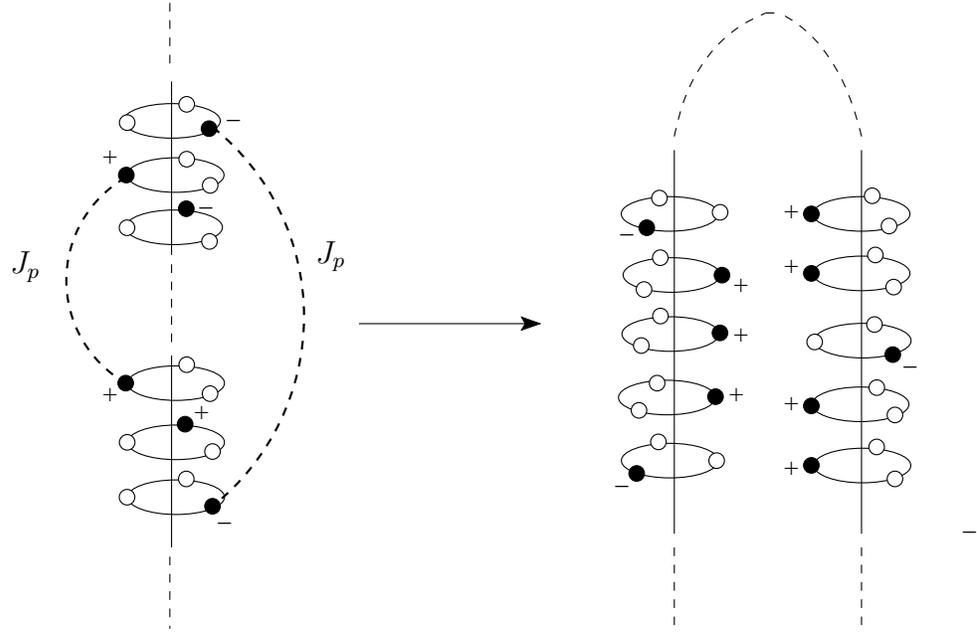}}
\put( 25, 34){$J_p$}
\put( 54, 35){$J_p$}
\end{picture}
\vspace*{2mm}
\caption{Illustration of polarity interactions.  Every pair $(i,j)$ of monomers
for which both $\xi_i=\xi_j$ (the two are of the same polarity,
denoted in the graph by `+' or `-') and $\phi_i=\phi_j$ (the two
are oriented towards the same side of the backbone) will give a reduction
of the total energy. The rationale is that such an arrangement
will make it easier for the polymer to fold into
an energetically favourable  conformation where hydrophobic monomers form the inner-residues (i.e. are shielded
from the solvent) and hydrophilic monomers
form the surface-residues (i.e. are exposed to the solvent).}
\label{fig:polarity}
\end{figure}
\begin{figure}[t]
\vspace*{-3mm}
\setlength{\unitlength}{1.4mm}
\begin{picture}(100,55)
\put( 30,  0){\epsfysize=50\unitlength\epsfbox{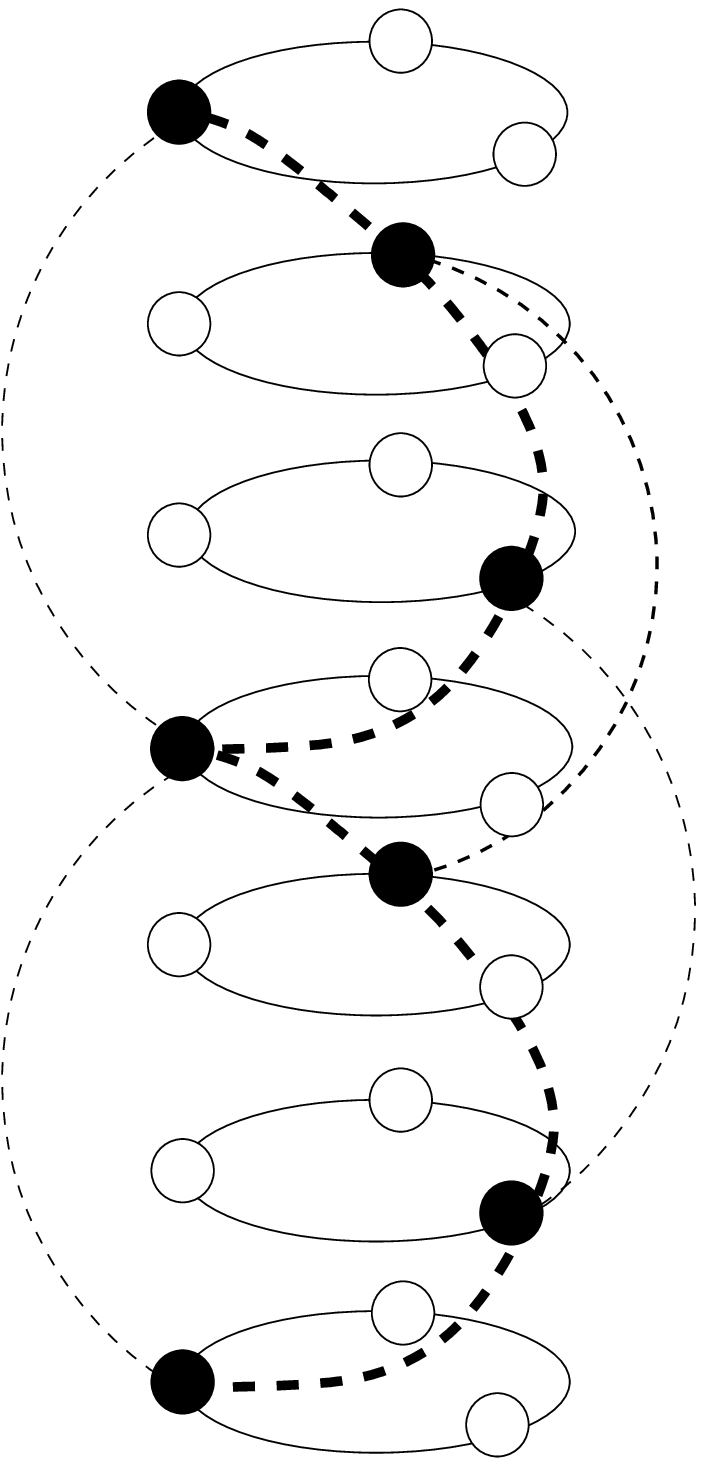}}
\put( 76, 15){\epsfysize=20\unitlength\epsfbox{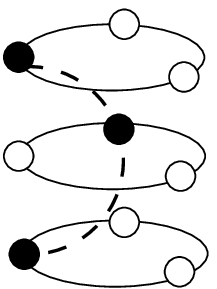}}
\put( 25, 32){$J_{Hb}^R$}
\put( 25, 15){$J_{Hb}^R$}
\put( 55, 30){$J_{Hb}^R$}
\put( 55, 18){$J_{Hb}^R$}
\put( 70, 17){$\lambda_{i-1}$}
\put( 70, 31){$\lambda_{i+1}$}
\put( 81, 24){$\lambda_{i}$}
\end{picture}
\vsp
\caption{Illustration of the hydrogen-bonding and steric energies. Left: hydrogen bonds of strength
$J_{Hb}^R$ are formed between monomers $i$ and $j$ whenever $|j-i|=q$, where $q$ represents the number of
available orientations ($q=3$ in this graph), and at the same time $\prod_{k=i}^{j-1}
\delta_{\phi_{k+1}-\phi_k,\frac{2\pi}{q}}=1$ (similarly for $J_{Hb}^L$).
The thick-dashed line in the left figure is a guide to the eye,
indicating the helical structure of the backbone induced by the
hydrogen bonds.
Right:
steric interactions impose
a specific preferred relative angle $a_i=(\phi_{i+1}-\phi_i)-
(\phi_i-\phi_{i-1})$, dependent on the (largely geometrical)
properties of the monomer type $\lambda_i$ present at site $i$.}
\label{fig:steric}
\end{figure}
\begin{enumerate}
\item
Polarity-induced energy (see figure \ref{fig:polarity}):
\begin{equation}
H_{\rm p}(\bphi)=-\frac{J_p}{N}\sum_{ij}\xi_i \xi_j~\delta_{\phi_i,\phi_j}
\label{eq:H_polar}
\end{equation}
This describes exchange-energies of monomer
pairs generated by their polarity type,
believed to be the main driving forces for compactification.
Proteins live in an aqueous environment, and amino-acids of
the same polarity prefer to co-align,
so that folding allows the chain to arrange for hydrophobic and
hydrophilic monomers to form the inner- and surface-residues of the
molecule, respectively. Equation (\ref{eq:H_polar}) describes this effect phenomenologically:
 $\xi_i$ indicates whether the monomer at site $i$ is
hydrophobic ($\xi_i=1$) or hydrophilic
($\xi_i=-1$), and we reduce
the configuration energy for every pair $(i,j)$ of monomer residues which are both of the same type and
which are also found in identical orientations relative to the backbone.
\item
hydrogen-bond energy (see figure \ref{fig:steric}):
\begin{equation}
H_{\rm Hb}(\bphi)=-\sum_i\left\{
J_{Hb}^L\!\prod_{k=0}^{q-1}\!\delta_{\phi_{i+k+1}-
 \phi_{i+k},{2\pi\over q}}
+
 J_{Hb}^R\!\prod_{k=0}^{q-1}\!\delta_{\phi_{i+k+1}
-\phi_{i+k},{- 2\pi\over q}}
\right\}
\label{eq:H_HB}
\end{equation}
The second contribution to the energy describes the effect of hydrogen bonding:
a monomer pair $(i,j)$ is coupled by a hydrogen bond
of strength $J_{Hb}^L$ or $J_{Hb}^R$ if and only if they are
spatially separated by exactly $q$ lattice sites  and if the relative angles
$\phi_{k+1}-\phi_k$ of all monomers $k=i,\ldots,i+q-1$ form a local helical
twist of $\pm\frac{2\pi}{q}$
(and therefore monomer $i$ and monomer $i+q$
have the same orientation relative to the backbone), such that intermediate monomers
do not block the formation of the
hydrogen bond.
\pagebreak
\item
Steric energy (see figure \ref{fig:steric}):
\begin{equation}
H_{\rm s}(\bphi)=-J_s\sum_{i}\cos[(\phi_{i+1}-\phi_i)-(\phi_i-\phi_{i-1})-a_i]
\label{eq:H_steric}
\end{equation}
This  describes local short-range steric monomer-monomer
interactions, favoring alignment of the relative angles
$(\phi_{i+1}-\phi_i)$ and $(\phi_i-\phi_{i-1})$ towards a specific preferred direction
$a_i$ which depends on the type of monomer present at site $i$.
\end{enumerate}

The various energy scales in the problem, and thus the relative
importance of the three types of forces, are controlled by the
non-negative coupling constants $\{J_p,J_{Hb}^{L,R},J_s\}$.
A preference for left-or right-handed helices can be built in by
modifying the balance between $J_{Hb}^L$ and $J_{Hb}^R$.
The quenched disorder in the problem is given by the realisation of the
(randomly drawn, but fixed) amino-acid sequence, i.e. the variables
$\{\xi_i,a_i\}$.
We denote the monomer type found at location $i$ in the chain by
$\lambda_i$, so that $\xi_i=\xi(\lambda_i)$ and
$a_i=a(\lambda_i)$. The disorder is characterized by the
distributions
\begin{eqnarray}
w[a,\xi]&=&\lim_{N\to\infty}\frac{1}{N}\sum_i\delta_{\xi,\xi_i}\delta[a-a_i]
=\sum_{\lambda}W(\lambda)~\delta_{\xi,\xi(\lambda)}\delta[a-a(\lambda)]
\label{eq:distributionofdisorder}
\\
W(\lambda)&=&\lim_{N\to \infty}\frac{1}{N}\sum_i\delta_{\lambda,\lambda_i}
\label{eq:distributionofacids}
\end{eqnarray}
Note that for random hetero-polymers the distribution $W(\lambda)$
is simply the {\em a priori} distribution according to which the monomers
were selected.
The marginal distribution specifying polarity statistics is
written as
\begin{equation}
w[\xi]=\int\!da~w[a,\xi]=\frac{1}{2}(1+p)\delta_{\xi,1}+\frac{1}{2}(1-p)\delta_{\xi,-1},
\label{eq:w_xi}
\end{equation}
with $p\in[-1,1]$.

We will solve our model in thermal equilibrium via a suitable combination of
mean- and random-field techniques \cite{bruinsma}, which will allow
us to evaluate the free energy per monomer $f$
in the thermodynamic limit:
\begin{equation}
f=-\lim_{N\to\infty}\frac{1}{\beta N}
\log \sum_{\bphi} e^{-\beta H(\bphi)}
\label{eq:free_energy_full}
\end{equation}
where $H(\bphi)=H_{\rm p}(\bphi)+H_{\rm Hb}(\bphi)+H_{\rm s}(\bphi)$.
The parameter $\beta$ is an effective
inverse temperature, which controls the amount of stochasticity in the underlying
dynamics (with $\beta=0$ and $\beta=\infty$ corresponding to purely random and purely deterministic
dynamics, respectively). The effective temperature will generally depend on various environmental factors, such as
solvent conditions.
We wish to emphasise that our present model takes into account the folding of the hetero-polymer only
as a general mechanism with which to realise the potential for energy gain via
polarity-induced forces, without specifying the detailed
three-dimensional
structure this reduction would give rise to.
It can consequently describe only  the formation of secondary structure as the result of
folding, not the emerging tertiary structure; this is the price to
be paid for exact analytical solvability.

Given the above definitions, it is natural to divide the monomers into two groups according to their
polarity, $\{1,\ldots,N\}=I_+\bigcup I_-$ with $I_\pm=\{i|~\xi_{i}=\pm 1\}$.
We note that $\lim_{N\to\infty}|I_\pm|/N=\frac{1}{2}(1\pm p)$.
Within each group one can define as natural observables
to measure the degree of polymer compactification (i.e. the impact
of the polarity-induced forces)
the distribution of monomer residue orientations, $P_+(\phi;\bphi)$ and
$P_-(\phi;\bphi)$:
\begin{equation}
P_\pm(\phi)=\lim_{N\to\infty}\bra
P_{\pm}(\phi;\bphi)\ket
~~~~~~~~
P_\pm(\phi;\bphi)=\frac{1}{|I_{\pm}|}\sum_{i\in I_\pm}
\delta_{\phi,\phi_i}
\label{eq:polarity}
\end{equation}
where $\bra\ldots\ket$ denotes an average over the Boltzmann distribution $p_\infty(\bphi)
\sim\exp[-\beta H(\bphi)]$.
Note that, by definition, $P_\pm(\phi)\in[0,1]$ and $\sum_\phi P_\pm(\phi)=1$. Note
also that due to the equivalence of all absolute orientations,
spontaneous symmetry breaking can occur.
In order to measure the degree of $L$(eft) or $R$(ight) chirality of the folded state, as induced by the steric
interactions and hydrogen bonds, we introduce the two order parameters
\begin{equation}
\chi_\pm=\lim_{N\to\infty}\bra \chi_{\pm}(\bphi)\ket
~~~~~~~~
\chi_\pm(\bphi)=\frac1N \sum_{i}
\prod_{k=0}^{m-1}\delta_{\phi_{i+k+1}-\phi_{i+k},\pm\frac{2\pi}{m}}
\label{eq:chirality}
\end{equation}
Thus $\chi_+=-\partial f/\partial J^L_{Hb}$ and
$\chi_-=-\partial f/\partial J^R_{Hb}$.

Before solving the full model it is instructive
to consider the various limiting cases one obtains by setting specific combinations
of the characteristic energies $\{J_s,J_p,J_{Hb}\}$ in (\ref{eq:H_polar}-\ref{eq:H_steric}) to zero.
First, in the absence of polarity interactions the model reduces to a one-dimensional random-field Potts
model with site-disorder, for which the free energy is known to be analytic for
finite temperatures, and there can be no phase transition. On the other
hand, in the absence of steric- and hydrogen-bond interactions the model reduces
to a mean-field model with site disorder. The most interesting scenario,
from a physical and a technical point of view, is the one where all three
forces are included. Due to the long-range interactions our model is expected to show
a phase transition, whereas the short-range interactions are
expected
to generate frustration phenomena such as hierarchies of discontinuous
transitions \cite{derridaetal} and non-analytic distribution functions for local
observables such as Devil's Staircases \cite{bruinsma}. An appealing feature of the model
is that, apart from the mean field forces, it is essentially one-dimensional
and thus allows for an exact solution
based on random-field techniques such as in
\cite{riera,brandtgross,bruinsma,rujan}.

\section{Solution of the Polarity Model \label{sec:polarity}}

In order to identify and interpret the properties of the full model, to be
analysed
in a subsequent section, we will now first solve our model in the absence of
short-range interactions, i.e. for $J_s=J_{Hb}^{L,R}=0$, so that
$H(\bphi)=H_{\rm p}(\bphi)$.

\subsection{Calculation of the Free Energy}

Upon using the simple identity $\sum_{ij}\delta_{\phi_i,\phi_j}=\sum_{\phi}\sum_{ij}
\delta_{\phi_i,\phi}\delta_{\phi_j,\phi}$ we can express the polarity
Hamiltonian (\ref{eq:H_polar}) in terms of the order parameters
(\ref{eq:polarity})
\begin{equation}
H_{\rm p}(\bphi)=-J_p N\sum_{\phi}
\left\{
\frac{|I_{+}|}{N} P_{+}(\phi;\bphi)
-
\frac{|I_{-}|}{N} P_{-}(\phi;\bphi)
\right\}^2
\end{equation}
Upon introducing delta functions to enforce the definitions (\ref{eq:polarity}), in integral representation,
we obtain the following expression for the free energy per site (\ref{eq:free_energy_full}):
\begin{equation}
f=-\lim_{N\to\infty}\frac{1}{\beta N}\log \int\prod_\phi\left[dP_\pm(\phi)
d\hat{P}_{\pm}
(\phi)\right]~e^{-N G\left[\{P_{\pm},\hat{P}_\pm\}\right]}
\label{eq:first_polar_free_energy}
\end{equation}
where
\begin{eqnarray*}
G\left[\{P_{\pm},\hat{P}_\pm\}\right]&=&-\frac{1}{4}\beta J_p
\sum_{\phi}\left\{(1+p)P_+(\phi)-(1-p)P_-(\phi)\right\}^2
\\
&&\hspace*{-10mm} -i\sum_{\phi}\left\{\hat{P}_+(\phi)P_+(\phi)+\hat{P}_-(\phi)P_-(\phi)\right\}
\\
&&\hspace*{-10mm}
-\frac{1}{2}(1+p)\log \sum_{\phi}e^{-2i\hat{P}_+(\phi)/(1+p)}
-\frac{1}{2}(1-p)\log \sum_{\phi}e^{-2i\hat{P}_-(\phi)/(1-p)}
\end{eqnarray*}
In the thermodynamic limit $N\to\infty$, the integral in (\ref{eq:first_polar_free_energy}) can be
evaluated via steepest descent. Derivation of $G[\ldots]$ with respect to $P_\pm(\phi)$
gives the equation
$i\hat{P}_\pm(\phi)=
\mp\frac{1}{2}(1\pm p)\beta J_p
\left[(1+p)P_+(\phi)-(1-p)P_-(\phi)\right]$, with which we
eliminate the conjugate order parameters.
This results in $f={\rm extr}_{\{L\}}f[\{L\}]$
\begin{equation}
f[\{L\}]=
\frac{J_p}{4} \sum_{\phi}L^2(\phi)
-\frac{1\!+\!p}{2\beta}
\log \sum_{\phi} e^{\beta J_p L(\phi)}
-\frac{1\!-\!p}{2\beta}
\log \sum_{\phi} e^{-\beta J_p L(\phi)}
\label{eq:polar_free_energy}
\end{equation}
where $L(\phi)=(1+p)P_+(\phi)-(1-p)P_-(\phi)$.
Extremisation with respect to the $L(\phi)$ leads
to a set of $q$ coupled saddle-point equations from which to solve
$\{L(\phi)\}$, in terms of which we can then also express our
original observables $P_{\pm}(\phi)$:
\begin{equation}
L(\phi)=(1\!+\!p)\frac{e^{ \beta J_p L(\phi)}}{\sum_{\phi'} e^{ \beta J_p L(\phi')}}
       -(1\!-\!p)\frac{e^{-\beta J_p L(\phi)}}{\sum_{\phi'} e^{-\beta J_p L(\phi')}}
\label{eq:fixed_point}
\end{equation}
\begin{equation}
P_\pm(\phi)=\frac{e^{\pm\beta J_p L(\phi)}}{\sum_{\phi'}e^{\pm\beta J_p
L(\phi')}}
\label{eq:SP_equations}
\end{equation}
Note that (\ref{eq:fixed_point}) is invariant under the transformation
$\{p,L(\phi)\}\to\{\!-\!p,\!-\!L(\phi)\}$ $\forall \phi$, and that $\sum_\phi L(\phi)=2p$.

The uniform
high temperature solution, where
 $L(\phi)=L^\star=2p/q$ for all $\phi$ and therefore $P_\pm(\phi)=\frac{1}{q}$ for all $\phi$,
 always satisfies (\ref{eq:fixed_point}).
Expansion of the free energy
(\ref{eq:polar_free_energy}) around the uniform solution $\{ L^*\}$ allows us to determine the
critical temperature $T_c=1/\beta_c$ where it becomes locally
unstable. For perturbations $\{\delta L\}$
orthogonal to $\{ L^*\}$, i.e. for which $\sum_{\phi}\delta L(\phi)=0$, we find
\begin{equation}
f[\{ L^\star\!+\delta L\}]= f[\{ L^\star\}]+
\frac{J_p^2}{2q}(\frac{q}{2 J_p}-\!\beta) \sum_{\phi}\delta^2
L(\phi)+\order(\delta^3 L)
\label{eq:perturbation}
\end{equation}
Hence a second-order phase transition to an ordered state takes place at
\begin{equation}
T_c=\beta^{-1}_{c}= \frac{2 J_p}{q}.
\label{eq:critical_beta}
\end{equation}
(or at a higher temperature, as a first-order transition).
This value is independent of the variable $p=\lim_{N\to \infty}\frac1N\sum_i \xi_i$, which measures the
balance between hydrophobic and hydrophilic monomers.

Similarly we can find the system's ground state, for any non-trivial value of
$m$. Let us define $L_g(\phi)=\lim_{T\to 0}L(\phi)$, $L_{+}=\max_\phi L_g(\phi)$ and
$L_{-}=\min_\phi L_g(\phi)$, and let us denote
the number of $\phi$ for which
$L_g(\phi)=L_+$ as $q_+\geq 1$ and the number for which
$L_g(\phi)=L_-$ as $q_-\geq 1$ (with $q_+ + q_-\leq q)$.
We assume $L_-<L_+$;
one can easily convince oneself that the alternative $L_-=L_+$, i.e. the high temperature
solution, will not give the ground state.
Taking the $T\to 0$ limit in the saddle-point equations (\ref{eq:fixed_point})
then shows that
$L_\pm=\frac{p\pm 1}{q_\pm}$, and that $L_g(\phi)=0$
for all $\phi$ such that $L_-<L_g(\phi)<L_+$. Thus
$L_g(\phi)$ can take only one of three different values.
The ground state energy per monomer, $u=\lim_{T\to 0}f$,
can subsequently be obtained as the $T\to 0$ limit of
(\ref{eq:polar_free_energy}):
\begin{eqnarray}
u&=&\frac{1}{2}J_p \min_{q_+,q_-}\left\{
\frac{1}{2} \sum_{\phi}L_g^2(\phi)
-(1\!+\!p)\max_\phi L_g(\phi)
+(1\!-\!p)\min_\phi L_g(\phi)\right\}
\nonumber \\
&=&-\frac{1}{4}J_p
\max_{q_+,q_-}\left\{
\frac{(1+p)^2}{q_+}
+\frac{(1-p)^2}{q_-}\right\}
=-\frac{1}{2}J_p(1+p^2)
\label{eq:groundstate_energy}
\end{eqnarray}
The minimum is obtained for $q_+=q_-=1$: there is one angle $\phi_+$ with $L_g(\phi_+)=p+1$,
there is one angle $\phi_-$
with $L_g(\phi_-)=p-1$, and the remaining $q-2$ orientations
have $L_g(\phi)=0$.  The ground state, written in terms of the monomer densities $P_\pm(\phi)$, is
\begin{eqnarray}
&P_+(\phi_+)=1, & ~~~~~~P_+(\phi)=0~~{\rm for~all}~~\phi\neq \phi_+
\label{eq:groundstate_plus}
\\
&P_-(\phi_-)=1, & ~~~~~~P_-(\phi)=0~~{\rm for~all}~~\phi\neq \phi_-
\label{eq:groundstate_minus}
\end{eqnarray}
All hydrophobic monomers cluster at some orientation $\phi_+$, and
all hydrophilic monomers cluster at a different orientation $\phi_-$, which
is indeed the energetically most favourable configuration for any value of $q$.
For $q>2$ this introduces a trivial degeneracy of the ground
state, since the choice made for $\phi_\pm$ is constrained only by
$\phi+\neq \phi_-$.

In general, non-trivial solutions of the non-linear fixed point
equations (\ref{eq:fixed_point}) can only be determined numerically, due to the
presence of two terms $\sum_\phi e^{\pm\beta J_p L(\phi)}$,
which act as normalisation constants for $P_\pm(\phi)$ and couple the $q$ equations in a transcendental manner.
However, for the two simplest scenarios $q=2$ (i.e. $\phi\in\{-\frac{\pi}{2},\frac{\pi}{2}\}$) and
$q=3$ (i.e. $\phi\in\{-\frac{2\pi}{3},0,\frac{2\pi}{3}\}$) it turns out that these terms can be
transformed away, and that an analytical solution is available.
We note that, due to the specific properties of the high temperature state
(where all $L(\phi)$ are identical) and of the ground state (where the $L(\phi)$
can take only one of three possible values), the $q>3$ phase
diagrams can at most differ quantitatively from that of the $q=3$
model (provided $q$ remains finite). We now solve our saddle-point equations (\ref{eq:fixed_point}) for
$q\in\{2,3\}$.

\subsection{Phase Diagram for $q=2$}

In the case where $q=2$ (two available orientations per monomer) we have
 $\phi\in\{-\frac{\pi}{2},\frac{\pi}{2}\}$, and we define $Z=\frac{1}{2}\beta
 J_p[L(\frac{1}{2}\pi)-L(-\frac{1}{2}\pi)]$. Since the two order parameters $L(\phi)$
also obey $\frac{1}{2}[L(\frac{1}{2}\pi)+L(-\frac{1}{2}\pi)]=p$,
one simply has
\[
L(\pm \frac{1}{2}\pi)=p\pm Z
\]
Insertion into (\ref{eq:fixed_point}) leads to a single Curie-Weiss equation
for $Z$:
\begin{equation}
Z=\tanh(\beta J_p Z)
\label{eq:curie_weiss}
\end{equation}
This predicts a second-order transition at
$\beta J_p=1$, in agreement with the critical temperature (\ref{eq:critical_beta})
for de-stabilization of the high-temperature solution found earlier.
The order parameter $Z$ is recognised to be simply the staggered
magnetisation $N^{-1}\sum_i\xi_i\sigma_i$ we would have generated if we had studied the $q=2$
model upon transforming $\phi_i=\frac{1}{2}\pi\sigma_i$, with $\sigma_i\in\{-1,1\}$
(this would have led to a Mattis-type \cite{mattis} Hamiltonian).
The order parameters $P_+(\phi)$ and $P_-(\phi)$ subsequently follow in terms of the solution
$Z$ of equation (\ref{eq:curie_weiss}) as
\begin{eqnarray*}
P_+(\frac{1}{2}\pi)=P_-(-\frac{1}{2}\pi)=\frac{1}{1+e^{-2\beta J_p
Z}}
\\
P_+(-\frac{1}{2}\pi)=P_-(\frac{1}{2}\pi)=\frac{1}{1+e^{2\beta J_p
Z}}
\end{eqnarray*}
For $T>T_c=J_p$ one simply recovers the uniform
state $P_+(\phi)=P_-(\phi)=\frac12$, for all $\phi$, as it should.
Below $T_c$ the system will choose to gradually align hydrophobic and
hydrophilic monomers and fold, with perfect alignment (or
separation) of the two polarity types at $T=0$.

\subsection{Phase Diagram for $q=3$}

In the case where $q=3$ (three possible orientations per monomer) we have
$\phi\in\{-\frac{2}{3}\pi,0,\frac{2}{3}\pi\}$. The possible
solutions of our saddle-point equation (\ref{eq:fixed_point}) can
be classified on the basis of the number of different values taken
by the three order parameters
$\{L(-\frac{2}{3}\pi),L(0),L(\frac{2}{3}\pi)\}$, as follows:
\begin{enumerate}
\item
All order parameters take the same value,
$L(-\frac{2}{3}\pi)=L(0)=L(\frac{2}{3}\pi)=\frac{2}{3}p$. This is
the uniform high temperature
state, which we have already encountered, and which according to (\ref{eq:critical_beta})
becomes locally unstable at $T_{c}=\frac{2}{3} J_p$.
\item
Exactly two order parameters take the same value. In view of the
invariance of equation (\ref{eq:fixed_point})
under permutations of the three allowed locations $\{-\frac{2}{3}\pi,0,\frac{2}{3}\pi\}$
we may without loss of generality put $L(\pm\frac{2}{3}\pi)=L_1$
and $L(0)=L_2$ (with $L_1\neq L_2$).
\item
All three order parameters are different:
$L(-\frac{2}{3}\pi)=L_1$, $L(0)=L_2$, $L(\frac{2}{3}\pi)=L_3$,
with $L_1\neq L_2\neq L_3$.
\end{enumerate}
We will show that, as the temperature is lowered,
first the type (ii) solution bifurcates
continuously from the type (i) solution at $T_c^I=\frac{2}{3}J_p$, and that the type (iii)
solution, in turn, bifurcates continuously from type (ii) at a lower temperature $T_c^{II}$.

 In order to find  the type (ii) solutions, and the critical temperature for which
these are created as bifurcations away from the uniform one,we introduce $Z=L_1-L_2$. Thus
\begin{eqnarray*}
L(\pm\frac{2}{3}\pi)&=&L_1=\frac{1}{3}(2p+Z)
\\
L(0)&=&L_2=\frac{1}{3}(2p-2Z)
\end{eqnarray*}
Insertion shows that such
states indeed solve
(\ref{eq:fixed_point}), with $Z$ following from
\begin{eqnarray}
Z=F(Z;\!\beta J_p)
\label{eq:fixed_typeb}\\
F(Z;\!K)=(1\!+ p)\frac{1\!-e^{-KZ}}{2+e^{-KZ}}
-
(1\!- p)\frac{1\!-e^{KZ}}{2+e^{KZ}}
\nonumber
\end{eqnarray}
The trivial solution $Z=0$ of (\ref{eq:fixed_typeb}) brings us back to the uniform state.
Bifurcations occur when $Z=F(Z;\beta J_p)$ and $1=\partial_Z F(Z;\beta
J_p)$; continuous bifurcations away from $Z=0$ occur when
$1=\lim_{Z\to 0}\partial_Z F(Z;\beta J_p)=\frac{2}{3}\beta J_p$.
This gives a second-order transition from state (i) to state (ii) at
the  critical temperature $T_c^{I}=\frac{2}{3}J_p$, i.e. precisely at
the point (\ref{eq:critical_beta}) where the uniform state was found to de-stabilise.
Since $\lim_{Z\to \pm\infty}F(Z;\!K)=\pm \frac{3}{2}-\frac{1}{2}p$ and $\lim_{Z\to 0}\partial_Z^2 F(Z;\!K)=-\frac{2}{9}K^2\leq 0$
there is no evidence for first-order transitions.

Next, in order to analyse the type (iii) solutions and to build in the normalisation
$\sum_{\phi}L(\phi)=2p$, we define $Z_1=L_1-L_2$ and
$Z_2=L_1-L_3$, such that
\begin{eqnarray*}
L(\frac{2}{3}\pi)&=&L_1=\frac{1}{3}(2p+Z_1+Z_2)
\\
L(0)&=&L_2=\frac{1}{3}(2p-2Z_1+Z_2)
\\
L(-\frac{2}{3}\pi)&=&L_3=\frac{1}{3}(2p+Z_1-2Z_2)
\end{eqnarray*}
This reduces our saddle-point equations (\ref{eq:fixed_point}) to
two coupled equations for $\{Z_1,Z_2\}$, which take the following form:
\begin{eqnarray}
Z_1=F(Z_1,Z_2;\!\beta J_p)~~~~~~~~Z_2=F(Z_2,Z_1;\!\beta J_p)
\label{eq:fixed_typec}
\\[1mm]
F(Z_1,Z_2;\!K)=
(1\!+p)\frac{1\!-e^{-K Z_1}}{1+e^{-K Z_1}+ e^{-K Z_2}}
-
(1\!-p)\frac{1\!-e^{K Z_1}}{1+e^{K Z_1}+e^{K Z_2}}
\nonumber
\end{eqnarray}
For $\{Z_1=0,Z_2\neq 0\}$ or $\{Z_2=0,Z_1\neq 0\}$ we return to a
state of type (ii), whereas the trivial solution $Z_1=Z_2=0$ brings us back to state
(i). Bifurcations occur when $(Z_1,Z_2)=\bF(Z_1,Z_2;\beta J_p)$ and
${\rm det}[\bone- (D\bF)(Z_1,Z_2)]=0$, where $\bF:\Re^2\to\Re^2$ denotes the non-linear mapping $(Z_1,Z_2)\to
(F(Z_1,Z_2;\!\beta J_p),F(Z_2,Z_1;\!\beta J_p))$
and $D\!\bF$ its Jacobian matrix. Thus, when the system is in a type (ii) state,
corresponding to e.g. $Z_1=Z$ and $Z_2=0$ with $Z$ given
as the solution of
(\ref{eq:fixed_typeb}), a continuous bifurcation is signaled by
\bd
{\rm det}\left|\begin{array}{cc}
1-(\partial_1F)(Z,0;\!\beta J_p)  & -(\partial_2 F)(Z,0;\!\beta J_p)\\
-(\partial_2 F)(0,Z;\!\beta J_p) & 1- (\partial_1F)(0,Z;\!\beta J_p)
\end{array}
\right|=0
\ed
Working out the partial derivatives shows that, since one of the off-diagonal terms vanishes,
this is equivalent to requiring either
\bd
\frac{1}{\beta J_p}=
\frac{1\!+p}{2+e^{-\beta J_p Z}}+\frac{1\!-p}{2+e^{\beta
J_p Z}}
\ed
or
\bd
\frac{1}{3\beta J_p}=
\frac{(1\!+p)e^{-\beta J_p Z}}{(2+e^{-\beta J_p Z})^2}
+\frac{(1\!-p)e^{\beta J_p Z}}{(2+e^{\beta
J_p Z})^2}
\ed
\begin{figure}[t]
\setlength{\unitlength}{1.2mm}
\vspace*{15mm}
\begin{picture}(100,55)
\put( 40,  0){\epsfysize=70\unitlength\epsfbox{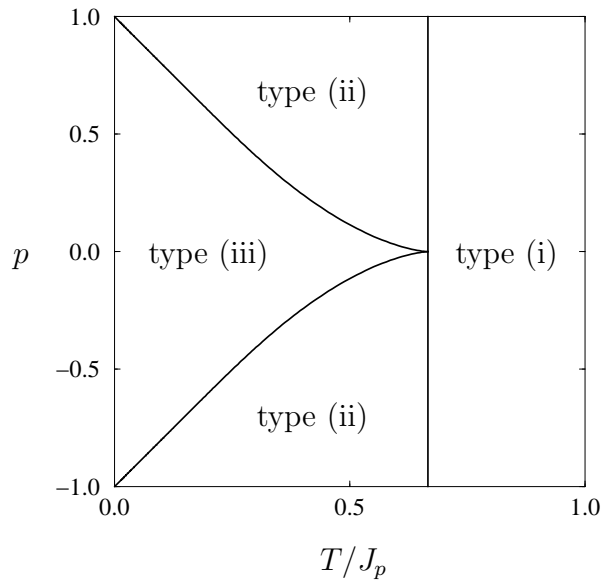}}
\put( 40, 36){$p$}
\put( 74,  2){$T/J_p$}
\put( 55, 36){type (iii)}
\put( 67, 54){type (ii)}
\put( 89, 36){type (i)}
\put( 67, 18){type (ii)}
\end{picture}
\caption{ Phase diagram of the polarity model for $q=3$, where
$\phi\in\{-\frac{2}{3}\pi,0,\frac{2}{3}\pi\}$.
Its regions are defined in terms of the number of different values
taken by the order parameters $\{L(\phi)\}$, and thus by the monomer distributions
$\{P_\pm(\phi)\}$,
at the three possible orientations:
(i) all three $L(\phi)$ are identical, (ii) only two of the $L(\phi)$ are identical,
(iii) all three $L(\phi)$ are different.
Within our model, these three types of phases, which  are separated by second order
transitions (indicated in the figure by solid
lines, with a tri-critical point at $(T/J_p,p)=(\frac{2}{3},0)$),
can be interpreted as representing different degrees of
folding.
Note that, in contrast to the case $q=2$, where
$\phi\in\{-\frac{1}{2}\pi,\frac{1}{2}\pi\}$, here the transitions do depend
on the polarity statistics as characterized by $p$.}
\label{fig:phase_diagram_polarity}
\end{figure}

\begin{figure}[t]
\vspace*{-5mm}
\setlength{\unitlength}{1.mm}
\begin{picture}(120,65)
\put( 16,  0){\epsfysize=140\unitlength\epsfbox{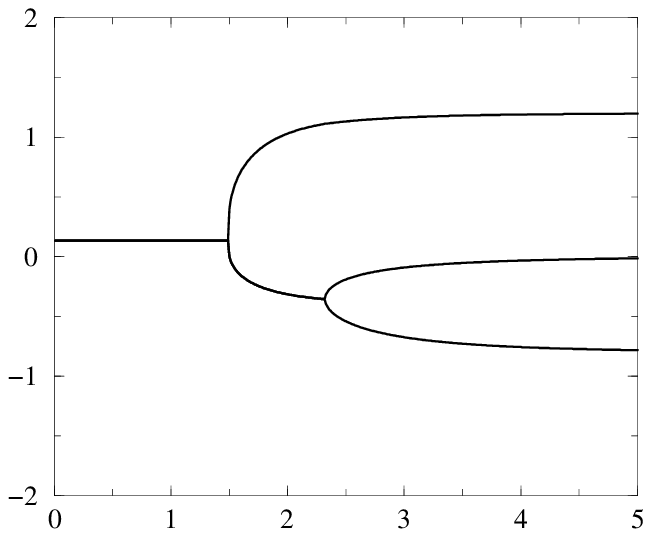}}
\put( 44, 55){$p=0.2$}
\put(  10, 27){$L(\phi)$}
\put( 47,  -1){$\beta J_p$}
\put( 95,  0){\epsfysize=140\unitlength\epsfbox{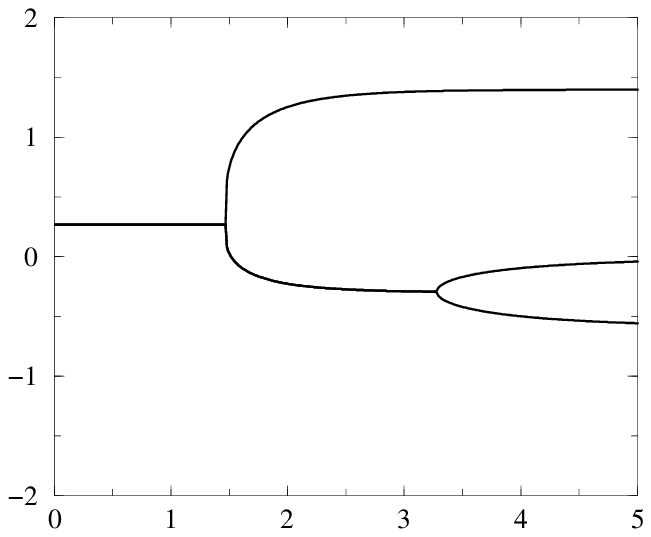}}
\put(123, 55){$p=0.4$}
\put( 89, 27){$L(\phi)$}
\put(126,  -1){$\beta J_p$}
\end{picture}
\vspace*{3mm}
\caption{The values taken by the three order parameters $L(\phi)$  for
the polarity model with $q=3$, i.e. $\phi\in\{-\frac{2}{3}\pi,0,\frac{2}{3}\pi\}$, as a function
of $\beta J_p$ (i.e. $J_p/T$) and for two different values of $p$.
They were obtained by
numerical solution of the saddle-point
equations (\ref{eq:fixed_point}).
The graphs shows the two phase transitions (i)$\to$(ii) and (ii)$\to$(iii) as continuous
bifurcations.
As predicted, the first transition occurs at $\beta J_p=\frac32$ (in both graphs),
whereas the location of the second transition depends on $p$. }
\label{fig:Ls}
\end{figure}
\noindent
The second equation signals a possible destabilisation within the
class of type (ii) solutions (which can only happen when there are multiple stable type (ii) solutions,
for which there is no evidence); the first equation describes the
creation/annihilation of type (iii) solutions from a type (ii) one. When solved in
combination with the saddle-point equation (\ref{eq:fixed_typeb}),
this latter equation gives the desired (second-order) (ii)$\to$(iii) transition line $T_c^{II}$.
The solution can be represented
conveniently  in the form of a parametrisation in the $(\beta J_p,p)$ plane, with $x=\beta J_p Z\in(-\infty,\infty)$:
\begin{equation}
\beta J(x)=\frac12\ \frac{\cosh (x) -1+x\sinh (x)}{\cosh (x)-1}
\label{eq:transition_line1}
\end{equation}
\begin{equation}
p(x)=\frac{x\cosh(x)+2x-3\sinh(x)}{1-\cosh(x)-x\sinh(x)}
\label{eq:transition_line2}
\end{equation}
Note that $\lim_{x\to\mp\infty}p(x)=\pm
1$ and that $\beta J(x)\sim \frac{1}{2}x$ as $x\to \infty$.
Equations (\ref{eq:transition_line1}-\ref{eq:transition_line2}) for the (ii)$\to$(iii) transition,
together with $\beta J_p=3/2$ (\ref{eq:critical_beta}) describing the (i)$\to$(ii)
transition, in fact represent all phase transitions in the $q=3$ system with polarity
energies only. This conjecture is based on extensive numerical exploration of the solutions of the fixed-point equations
(\ref{eq:fixed_point}).

In figure \ref{fig:phase_diagram_polarity} we show the resultant phase diagram of the
polarity model for $q=3$, i.e. for $\phi\in\left\{-\frac{2\pi}{3},0,\frac{2\pi}{3}\right\}$, in
the $(T/J_p,p)$ plane. It consists of regions characterised by the number of different values taken
by the three order parameters $\{L(\phi)\}$, and therefore by the monomer distributions $\{P_\pm(\phi)\}$
at the three possible
orientations. All regions are separated by
second-order transition lines, viz. (\ref{eq:critical_beta}) and
(\ref{eq:transition_line1}-\ref{eq:transition_line2}).
For $T/J_p>\frac{3}{2}$ (the high-temperature region) the only possible
solution of (\ref{eq:fixed_point}), for any $p$, is $L(\phi)=\frac{2}{3}p$ for all
$\phi$; here the monomers have no preferred orientation (to be interpreted as resulting in a swollen state).
For $T/J_p\leq \frac{2}{3}$ the equilibrium solution will depend
on the value of the polarity statistics parameter $p$. In region
(ii) the monomers exhibit some degree of orientation preference
(to be interpreted as resulting in a partially folded state),
whereas in region (iii) one finds a highly orientation specific
solution (to be interpreted as resulting in a fully folded state).
Note that, in view of the fact that also for $q>3$ the system will
in equilibrium allow for at most three different values for the
order parameters $L(\phi)$, see
(\ref{eq:groundstate_plus},\ref{eq:groundstate_minus}),
one must expect the $q>3$ phase diagrams to be
qualitatively similar to the $q=3$ one, with only $q$-dependent
re-scaling and weak deformations of transition lines.

In figure \ref{fig:Ls} we show the values of the three order parameters $L(\phi)$, from
which the monomer densities $P_\pm(\phi)$ follow via (\ref{eq:SP_equations}),
as a function of $\beta J_p$, for $p=0.2$ (left graph) and $p=0.4$ (right graph).
These values are obtained by numerical solution of the
saddle-point equations (\ref{eq:fixed_point}).
We clearly observe the point where type (ii) solutions (two possible values for the $L(\phi)$)
bifurcate from the type (i) solution (all $L(\phi)$ are identical), at
$\beta J_p=3/2$ for both graphs. In contrast, the location of the second
bifurcation from type (ii) to type (iii) solutions is indeed seen to depend on
the parameter $p$, as predicted.
We also observe how for $\beta \to \infty$ the system approaches the
ground state
(\ref{eq:groundstate_plus},\ref{eq:groundstate_minus}),
where $L(\phi)\in\{p-1,0,p+1\}$.

\section{Solution of the Full Model for $q=2$}

We will now turn to the full model described by the combination of all
three energy contributions (\ref{eq:H_polar}-\ref{eq:H_steric}).
Since we now have a Hamiltonian with both (site) disorder and short-range
interactions, a simple mean-field approach such as that used in
the previous section will no longer apply. Here our solution will
be based on a suitable adaptation of the random-field
techniques of \cite{brandtgross,bruinsma}.
We will, for simplicity, consider only the simplest non-trivial case $q=2$,
where $\phi_i=\frac{1}{2}\pi \s_i$ with $\s_i\in\{-1,1\}$.
Our orientation variables can now be replaced by Ising spins, which leads
to significant simplifications. For instance,
the various terms in the Hamiltonian reduce to (upon dropping the
irrelevant constants):
\begin{eqnarray}
H_{\rm p}(\bsigma)=-\frac{J_p}{2N}\sum_{ij}\sigma_i\xi_i \xi_j\sigma_j
\label{eq:Ising_polar}
\\
H_{\rm Hb}(\bsigma)=-\frac{1}{2}J_{Hb}\sum_i
[1-\s_{i}\s_{i+1}][1-\s_{i}\s_{i-1}]
\label{eq:Ising_HB}
\\
H_{\rm s}(\bsigma)=-J_s\sum_{i}\eta_i~\sigma_{i+1}\sigma_{i-1}
\label{eq:Ising_steric}
\end{eqnarray}
with $J_{Hb}=\frac{1}{2}(J_{Hb}^L+J_{Hb}^R)$, and with
$\eta_i=\cos[a_i]$.
Left- and right chirality energies have become identical, as
expected for $\phi_i\in\{-\frac{1}{2}\pi,\frac{1}{2}\pi\}$.
The two `chirality' order parameters (\ref{eq:chirality}) reduce to
\begin{equation}
\chi=\lim_{N\to\infty}\bra \chi(\bsigma)\ket
~~~~~~~~
\chi(\bsigma)=
\frac{1}{4N}\sum_{i}[1-\s_{i}\s_{i+1}][1-\s_{i}\s_{i-1}]
\label{eq:chirality_ising}
\end{equation}
The joint distribution $\tilde{w}[\eta,\xi]$ of the disorder variables $\{\eta_i,\xi_i\}$
(which are independent for different sites) follows from
(\ref{eq:distributionofdisorder}):
\begin{equation}
\tilde{w}[\eta,\xi]
=\sum_{\lambda}W(\lambda)~\delta_{\xi,\xi(\lambda)}\delta\left[\eta-\cos[a(\lambda)]\right]
\label{eq:Ising_disorder}
\end{equation}

\subsection{Calculation of the Free Energy}

We note that the polarity energy (\ref{eq:Ising_polar})
can be written in terms of the
`staggered magnetisation'
\begin{equation}
m(\bsigma)=\frac{1}{N}\sum_i\xi_i\sigma_i
\label{eq:polarity_ising}
\end{equation}
in the Mattis \cite{mattis} form $H_{\rm p}(\bsigma)=-\frac{1}{2}J_p Nm^2(\bsigma)$.
We isolate the order parameter $m$ in the expression for the free
energy per site (\ref{eq:free_energy_full}), by inserting $1=\int\!dm~\delta[m-\frac1N
\sum_i \s_i\xi_i]$. Writing the delta function in
integral representation then leads to
\begin{equation}
f=-\lim_{N\to\infty}\frac{1}{\beta N}\log\int dm d\hat{m}\
e^{-\beta N G_N(m,\hat{m})}
\label{eq:free_energy_integral}
\end{equation}
\begin{equation}
G_N(m,\hat{m})=-im\hat{m}-\frac12 J_p m^2-\frac{1}{\beta N}\log Z_N(-i\beta\hat{m})
\label{eq:free_energy_precursor}
\end{equation}
where the complicated (short-range) part of the partition sum has now been concentrated in
the function $Z_N(x)$:
\begin{equation}
Z_{N}(x)=\!\sum_{\s_1\ldots\s_N}e^{\frac{1}{2}\beta J_{Hb}\sum_i
[1-\s_i\s_{i+1}][1-\s_i\s_{i-1}]+\beta J_s\sum_{i}\s_{i-1}\eta_i\s_{i+1}
+x \sum_{i}\s_i\xi_i}
\label{eq:partition_sum}
\end{equation}
(with $\sigma_0=\sigma_{N+1}\equiv 0$).
The integral in (\ref{eq:free_energy_integral}) can for $N\to\infty$ be evaluated
via steepest descent, and will be dominated by
the saddle points of the exponent $G_\infty(m,\hat{m})$.
After elimination of $\hat{m}$ via the saddle-point equation $i\hat{m}=-J_p m$,
we can thus write the asymptotic free energy per monomer
(\ref{eq:free_energy_precursor}) as
\begin{equation}
f={\rm extr}_{m}\left\{\frac12 J_p m^2-\lim_{N\to\infty}\frac{1}{\beta N}\log
Z_N(\beta J_p m)\right\}
\label{eq:free_energy}
\end{equation}
In order to
calculate the partition sum (\ref{eq:partition_sum}) we will
employ the random-field techniques of \cite{brandtgross,bruinsma}.  We
condition the function $Z_N(x)$ on the values $\{\s_{N-1},\s_N\}$ of the two spins at the end
of the chain:
\begin{equation}
\hspace*{-18mm}
Z_{\sigma \sigma^\prime}^{(N)}(x)
=\!\sum_{\s_1\ldots\s_N}e^{\frac{1}{2} \beta J_{Hb}\sum_i
[1-\s_i\s_{i+1}][1-\s_i\s_{i-1}]+J_s\sum_{i}\s_{i-1}\eta_i\s_{i+1}
+x \sum_{i}\s_i\xi_i}~\delta_{\s_{N-1},\sigma}\delta_{\s_N,\sigma^\prime}
\label{eq:conditioned_def}
\end{equation}
with $Z_N(x)=\sum_{\sigma \sigma^\prime=\pm1}Z_{\sigma, \sigma^\prime}^{(N)}(x)$.
The addition of an
extra monomer to the chain, i.e. $N\to N+1$, then leads to the
following recurrent relation for the conditioned partition
functions:
\begin{equation}
\left(\begin{array}{c}
Z^{(N+1)}_{++}(x) \\ Z^{(N+1)}_{+-}(x) \\ Z^{(N+1)}_{-+}(x) \\
Z^{(N+1)}_{--}(x)
\end{array}\right)
=
\bM_{N+1}(x) \bT_{N}
\left(\begin{array}{c}
Z^{(N)}_{++}(x) \\ Z^{(N)}_{+-}(x) \\ Z^{(N)}_{-+}(x) \\
Z^{(N)}_{--}(x)
\end{array}\right)
\label{eq:recurrence_rel}
\end{equation}
in which the 4$\times$4 matrices $\bM_i(x)$ and $\bT_i$ are defined
as
\begin{equation}
\bM_{i}(x)=
\left(\begin{array}{cccc}
e^{x \xi_{i}} & 0 & 0 & 0 \\
0 & e^{-x \xi_{i}} & 0 & 0 \\
0 & 0 & e^{x \xi_{i}} & 0 \\
0 & 0 & 0 & e^{-x \xi_{i}}
\end{array}\right)
\label{eq:M_matrix}
\end{equation}
\begin{equation}
\bT_{i}=
\left(\begin{array}{cccc}
e^{\beta J_s \eta_i} & 0 & e^{-\beta J_s \eta_i -\beta
J_{Hb}} & 0 \\
e^{-\beta J_s \eta_i +\beta J_{Hb}} & 0 & e^{\beta J_s \eta_i +2\beta
J_{Hb}} & 0 \\
0 & e^{\beta J_s \eta_i + 2 \beta J_{Hb}} & 0 & e^{-\beta J_s
\eta_i+\beta J_{Hb}} \\
0 & e^{-\beta J_s \eta_i - \beta J_{Hb}} & 0 & e^{\beta J_s \eta_i}
\end{array}\right)
\label{eq:T_matrix}
\end{equation}
As a result we can now write the short-range partition sum $Z_{N}(x)$ (\ref{eq:partition_sum})
in terms of the random matrices (\ref{eq:M_matrix},\ref{eq:T_matrix}),
where the randomness is in the $\{\xi_i,\eta_i\}$,  as
\begin{equation}
Z_N(x)
=
\left(\begin{array}{c}
1 \\ 1 \\ 1 \\ 1
\end{array}\right)
\cdot
\left[\prod_{i=3}^{N} \bM_{i+1}(x) \bT_i \right]
\left(\begin{array}{c}
Z^{(2)}_{++}(x) \\ Z^{(2)}_{+-}(x) \\ Z^{(2)}_{-+}(x) \\
Z^{(2)}_{--}(x)
\end{array}\right)
\end{equation}
The (random) matrix product will be evaluated in terms of
the following (non-negative) stochastic quantities, which represent the different ratios of the
conditioned partition sums (\ref{eq:conditioned_def}):
\begin{equation}
k_{j}^{(1)}=e^{-2x \xi_{j}} \frac{Z_{++}^{(j)}}{Z_{+-}^{(j)}}
\hspace{10mm}
k_j^{(2)}=e^{2x \xi_{j}}\frac{Z_{+-}^{(j)}}{Z_{-+}^{(j)}}
\hspace{10mm}
k_j^{(3)}=e^{-2x \xi_{j}}\frac{Z_{-+}^{(j)}}{Z_{--}^{(j)}}
\label{eq:ratios_def}
\end{equation}
From the recurrence relation (\ref{eq:recurrence_rel}) it follows
that the variables $k_j^{(\ell)}$ are, in turn, generated by iteration of the following mapping:
\begin{equation}
k_{j+1}^{(1)}=\frac{e^{\beta J_s \eta_{j}} k_{j}^{(1)}k_{j}^{(2)}+
e^{-\beta J_s\eta_{j}-\beta J_{Hb}}}
{e^{-\beta J_s \eta_{j}} k_{j}^{(1)}k_{j}^{(2)}+
e^{\beta J_s\eta_{j}+\beta J_{Hb}}}~
e^{-\beta J_{Hb}}
\label{eq:k1}
\end{equation}
\begin{equation}
k_{j+1}^{(2)}=\frac{e^{-\beta J_s\eta_j}k_j^{(1)}k_j^{(2)}+e^{\beta J_s \eta_j+\beta
J_{Hb}}}{e^{\beta J_s\eta_j +\beta J_{Hb}}k_j^{(2)}k_j^{(3)}+e^{-\beta J_s \eta_j}}
~k_{j}^{(3)} e^{2x \xi_{j}}
\label{eq:k2}
\end{equation}
\begin{equation}
k_{j+1}^{(3)}=\frac{e^{\beta J_s \eta_{j}+\beta J_{Hb}} k_{j}^{(2)}k_{j}^{(3)}+
e^{-\beta J_s\eta_{j}}}
{e^{-\beta J_s \eta_{j}-\beta J_{Hb}} k_{j}^{(2)}k_{j}^{(3)}+
e^{\beta J_s\eta_{j}}}~ e^{\beta J_{Hb}}
\label{eq:k3}
\end{equation}
We now use
\begin{equation}
\frac{1}{\beta N}\log Z_N(x )=\frac{1}{\beta N}\log Z^{(N)}_{--}(x ) +\mc{O}(\frac1N)
\end{equation}
and work out the conditioned partition function $Z^{(N)}_{--}(x )$ iteratively,
via the the recurrence relation
(\ref{eq:recurrence_rel}):
\begin{eqnarray}
\frac{1}{N}\log Z^{(N)}_{--}(x )
&=&
\frac{1}{N}\log Z^{(N-1)}_{--}(x )
-\frac{x  \xi_{N}}{ N}
\nonumber \\
&&+\frac{1}{N}\log \left\{
e^{-\beta J_s \eta_{N-1} - \beta J_{Hb}}
k_{N-1}^{(2)}k_{N-1}^{(3)}
+e^{\beta J_s \eta_{N-1}}
\right\}
\end{eqnarray}
Further iteration of this relation gives
\begin{equation}
\hspace*{-15mm}
\lim_{N\to\infty}\frac{1}{N}\log Z^{(N)}_{--}(x )
=\int\!d\bk d\eta~P(\bk,\eta)
\log \left\{
e^{-\beta J_s \eta - \beta J_{Hb}}
k^{(2)} k^{(3)}
+e^{\beta J_s \eta}
\right\}
-x  p
\label{eq:iterated}
\end{equation}
with $\bk=(k^{(1)},k^{(2)},k^{(3)})$, where
$p=\int\!d\eta\sum_{\xi}\xi\tilde{w}[\eta,\xi]$
(see equation (\ref{eq:Ising_disorder})), and with
\begin{equation}
P(\bk,\eta)=\lim_{N\to\infty}\frac{1}{N}\sum_i\delta[\eta-\eta_i]\delta[\bk-\bk_i]
\end{equation}
Provided the stochastic process (\ref{eq:k1}-\ref{eq:k3})
is ergodic, the distribution $P(\bk,\eta)$ will be identical to
the (joint) stationary distribution of the pair $\{\bk,\eta\}$,
i.e. we may write
$P(\bk,\eta)=\lim_{N\to\infty}\frac{1}{N}\sum_i \bra\delta[\eta-\eta_i]\delta[\bk-\bk_i]\ket$.
Since $\bk_{i}$ is always statistically independent of $\eta_{i}$
according to
(\ref{eq:k1}-\ref{eq:k3}) ($\bk_i$ depends only on those $\eta_j$ and $\xi_j$ with $j<i$),
we have $\bra \delta[\eta-\eta_i]\delta[\bk-\bk_i]\ket=
\bra\delta[\eta-\eta_i]\ket\bra \delta[\bk-\bk_i]\ket$.
Hence $P(\bk,\eta)=P_\infty(\bk|x )\tilde{w}[\eta]$, where $P_\infty(\bk|x )$
is the invariant distribution of the process
(\ref{eq:k1}-\ref{eq:k3}) (which is parametrised by $x $, due to the occurrence of $x $ in (\ref{eq:k2})) and
where $\tilde{w}[\eta]=\sum_\xi \tilde{w}[\eta,\xi]$.
We thereby find (\ref{eq:iterated}) being replaced by
\begin{eqnarray}
\lim_{N\to\infty}\frac{1}{N}\log Z^{(N)}_{--}(x )
&=& -x  p
\nonumber \\
&&\hspace*{-25mm}
+\int\!d\bk~P_\infty(\bk|x )\int\!d\eta~ \tilde{w}[\eta]
\log \left\{
e^{-\beta J_s \eta - \beta J_{Hb}}
k^{(2)} k^{(3)}
+e^{\beta J_s \eta}
\right\}
\label{eq:iterated_simpler}
\end{eqnarray}
As a final consequence  we can now write the free energy per monomer (\ref{eq:free_energy}) as
\begin{eqnarray}
f={\rm extr}_{m}\left\{\frac12 J_p m^2+ J_p m p
\right.
\nonumber \\
&&
\hspace*{-45mm}
\left.
-\frac{1}{\beta}\int\!d\bk~P_\infty(\bk|\beta J_p m)\int\!d\eta~ \tilde{w}[\eta]
\log \left[
e^{-\beta J_s \eta - \beta J_{Hb}} k^{(2)} k^{(3)}
+e^{\beta J_s \eta}
\right]
\right\}
\label{eq:free_energy_final}
\end{eqnarray}
where the invariant measure $P_\infty(\bk|x )$ of the process (\ref{eq:k1}-\ref{eq:k3})
is to be solved from
\begin{equation}
P_\infty(\bk|x )=\int\!d\bk^\prime~P_\infty(\bk^\prime|x )
\int\!d\eta\sum_\xi
~\tilde{w}[\eta,\xi]~
\delta\left[\bk-{\cal F}(\bk^\prime|x ,\eta,\xi)\right]
\end{equation}
with
\begin{equation}
\left(\!\begin{array}{c} {\cal F}_1(\bk|x ,\eta,\xi)\\[2mm] {\cal
F}_2(\bk|x ,\eta,\xi)\\[2mm]
{\cal F}_3(\bk|x ,\eta,\xi)\end{array}\!\right)=
\left(\!\begin{array}{c}
\frac{e^{\beta J_s \eta} k_1 k_2+e^{-\beta J_s\eta-\beta J_{Hb}}}
{e^{-\beta J_s \eta} k_1 k_2+e^{\beta J_s\eta+\beta J_{Hb}}}~e^{-\beta J_{Hb}}
\\[2mm]
\frac{e^{-\beta J_s\eta} k_1 k_2+e^{\beta J_s \eta+\beta J_{Hb}}}
{e^{\beta J_s\eta +\beta J_{Hb}}k_2 k_3 +e^{-\beta J_s \eta}}
~k_3 e^{2x \xi}
\\[2mm]
\frac{e^{\beta J_s \eta +\beta J_{Hb}} k_2 k_3 + e^{-\beta J_s\eta}}
{e^{-\beta J_s \eta-\beta J_{Hb}} k_2 k_3 +
e^{\beta J_s\eta}}~ e^{\beta J_{Hb}}
\end{array}\!\right)
\label{eq:distribution_def}
\end{equation}
In the case of the one-dimensional random-field Ising model
\cite{bruinsma,aeppli}, for which the analysis is very similar, the corresponding distribution
$P_\infty(\bk)$ is known,  at least in
certain parameter regions, to become  highly non-trivial and acquire the form
of the derivative of a Devil's Staircase.  To our knowledge,
no general analytic expression has been
derived to describe $P_\infty(\bk)$ for finite temperatures. Nevertheless, for
the purpose of the present paper it is only a simple numerical exercise to
evaluate $P_\infty(\bk|x )$ directly by iteration of
(\ref{eq:distribution_def}), for values of $\{\eta,\xi\}$ drawn
randomly according to $\tilde{w}[\eta,\xi]$.

\subsection{Simple Limiting Cases}

Before converting our general results into phase diagrams
we will first carry out benchmark tests of our expressions, by inspecting
simple limits.
\begin{itemize}
\item
Firstly, in the absence of
short-range interactions, i.e. for $J_{Hb}=J_s=0$, the expression for the asymptotic free
energy per site (\ref{eq:free_energy_final}) should reduce to
the $q=2$ version of (\ref{eq:polar_free_energy}), which ought to be simply the free
energy of the infinite-range Mattis model \cite{mattis}. Indeed, we find that for
$J_s=J_{Hb}=0$ the
mapping (\ref{eq:distribution_def}) reduces to
\begin{equation}
\left(\!\begin{array}{c} {\cal F}_1(\bk|x ,\eta,\xi)\\[0mm] {\cal
F}_2(\bk|x ,\eta,\xi)\\[0mm]
{\cal F}_3(\bk|x ,\eta,\xi)\end{array}\!\right)=
\left(\!\begin{array}{c} 1 \\[0mm]
\frac{k_1 k_2+1}
{k_2 k_3 +1}~k_3 e^{2x \xi}
\\[0mm]
1
\end{array}\!\right)
\end{equation}
Hence $P_\infty(\bk|x )=\delta[k_1-1]\delta[k_3-1]P_\infty(k_2|x )$,
with
\begin{equation}
P_\infty(k_2|x )=
\frac{1}{2}(1+p)
\delta[k_2- e^{2x }]
+\frac{1}{2}(1-p)\delta[k_2- e^{-2x }]
\label{eq:distr_simple}
\end{equation}
Substitution into (\ref{eq:free_energy_final}), for $J_{Hb}=J_s=0$, gives
\begin{eqnarray}
f={\rm extr}_{m}\left\{\frac12 J_p m^2
-\frac{1}{\beta }\log 2\cosh(\beta J_p m)
\right\}
\end{eqnarray}
which is indeed the well-known asymptotic free energy per site of an
infinite-range Mattis magnet \cite{mattis}.
\item
Secondly, for $J_{Hb}=0$ and $\eta_i=\xi_{i}=1$ for all $i$
(i.e. $\tilde{w}[\eta,\xi]=\delta[\eta-1]\delta[\xi-1]$ and $p=1$)
the macroscopic laws of our model should reduce to those of
the \cite{hebbian}, which describes pattern recall in recurrent neural
networks with competition between short-range and long-range
information processing, for the simplest `one-pattern' scenario.
For $J_{Hb}=0$ and $\tilde{w}[\eta,\xi]=\delta[\eta-1]\delta[\xi-1]$ the mapping
(\ref{eq:distribution_def}) becomes fully deterministic, and takes
the form
\begin{equation}
\left(\!\begin{array}{c} {\cal F}_1(\bk|x )\\[2mm] {\cal
F}_2(\bk|x )\\[2mm]
{\cal F}_3(\bk|x )\end{array}\!\right)=
\left(\!\begin{array}{c}
\frac{e^{\beta J_s} k_1 k_2+e^{-\beta J_s}}
{e^{-\beta J_s} k_1 k_2+e^{\beta J_s}}
\\[2mm]
\frac{e^{-\beta J_s} k_1 k_2+e^{\beta J_s}}
{e^{\beta J_s}k_2 k_3 +e^{-\beta J_s}}
~k_3 e^{2x }
\\[2mm]
\frac{e^{\beta J_s } k_2 k_3 + e^{-\beta J_s}}
{e^{-\beta J_s } k_2 k_3 +
e^{\beta J_s}}
\end{array}\!\right)
\label{eq:deterministic_map}
\end{equation}
and $P_\infty(\bk|x )=\delta[\bk-\bk^\star(x )]$, where $\bk^\star(x )$
denotes the fixed-point of the mapping
(\ref{eq:deterministic_map}) with non-negative components, which (in line with our previous assumption
of ergodicity of the original process (\ref{eq:k1}-\ref{eq:k3}))
we assume to be unique.
We observe that (\ref{eq:deterministic_map}) preserves
$k_1=k_3$,
 and the remaining components of the fixed-point $\bk^\star(x )=(k^\star_1,k^\star_2,k^\star_1)$ must
obey
\begin{equation}
\left(\!\begin{array}{c} k^\star_1\\[2mm] k^\star_2\end{array}\!\right)=
\left(\!\begin{array}{c}
\frac{e^{\beta J_s} k^\star_1 k^\star_2+e^{-\beta J_s}}
{e^{-\beta J_s} k^\star_1 k^\star_2+e^{\beta J_s}}
\\[2mm]
\frac{e^{-\beta J_s} k^\star_1 k^\star_2+e^{\beta J_s}}
{e^{\beta J_s}k^\star_2 k^\star_1 +e^{-\beta J_s}}
~k^\star_1 e^{2x }
\end{array}\!\right)
\end{equation}
This (in turn)
gives $(k_1^\star,k_2^\star)=(k^\star,e^{2x })$, where (upon substituting
$x =\beta J_p m$) $k^\star$
is the non-negative solution of
\begin{equation}
k^\star=
\frac{e^{\beta (J_s+J_pm)} k^\star +e^{-\beta (J_s+J_pm)}}
{e^{-\beta (J_s-J_p m)} k^\star +e^{\beta (J_s-J_p m)}}
\label{eq:equation_kstar}
\end{equation}
Insertion into (\ref{eq:free_energy_final}) gives us
\begin{equation}
f={\rm extr}_{m}\left\{\frac12 J_p m^2
-\frac{1}{\beta}
\log \left[e^{-\beta (J_s- J_p m)}  k^\star +e^{\beta (J_s- J_p m)}
\right]
\right\}
\label{eq:solution_previous}
\end{equation}
It follows from (\ref{eq:equation_kstar}) that the quantity
$\lambda=e^{-\beta (J_s- J_p m)}  k^\star +e^{\beta (J_s- J_p m)}$
occurring in (\ref{eq:solution_previous}) obeys
$(\lambda-e^{\beta(J_2+J_p m)})(\lambda-e^{\beta(J_s-J_p m)})=e^{-2\beta
J_s}$, which we recognise  as the eigenvalue equation of the
transfer matrix
\begin{equation}
\bT=
\left(\begin{array}{cc}
e^{\beta (J_s+ J_p m)} & e^{-\beta J_s} \\
e^{-\beta J_s} & e^{\beta (J_s - J_p m)}
\end{array}\right)
\end{equation}
This shows that the free energy (\ref{eq:solution_previous}) is
indeed identical to that of \cite{hebbian}.
\end{itemize}

\subsection{Phase Diagrams and Comparison with Numerical Experiments}

In order to obtain phase diagrams we finally have to calculate the local extrema of
a free energy surface $f[m]$, the argument of the extremisation in (\ref{eq:free_energy_final}),
which still depends on the choice made for the
statistics of the monomer properties $\{\xi,\eta\}$. Here we apply our theory
to the simple example
 $\tilde{w}[\eta,\xi]=\frac{1}{4}[\delta(\eta+1)+\delta(\eta-1)][\delta(\xi+1)+\delta(\xi-1)]$,
hence also $p=0$.
In this case the free energy surface $f[m]$ simplifies to
\begin{eqnarray}
f[m]&=&\frac12 J_p m^2-\frac{1}{2\beta}\int\!d\bk~P_\infty(\bk|\beta J_p m)
\nonumber \\
&&
\times
\log\left[
(e^{-\beta (J_s +  J_{Hb})} k_2 k_3+e^{\beta J_s})
(e^{\beta (J_s - J_{Hb})} k_2 k_3+e^{-\beta J_s})
\right]
\label{eq:free_energy_examples}
\end{eqnarray}
where the invariant measure $P_\infty(\bk|x )$ of the process (\ref{eq:k1}-\ref{eq:k3})
is to be solved from
\begin{equation}
P_\infty(\bk|x )=\frac{1}{4}\int\!d\bk^\prime~P_\infty(\bk^\prime|x )
\sum_{\eta=\pm 1}\sum_{\xi=\pm 1}
\delta\left[\bk-{\cal F}(\bk^\prime|x ,\eta,\xi)\right]
\label{eq:stationary_examples}
\end{equation}
with the mapping defined in (\ref{eq:distribution_def}).
We determine the solution of (\ref{eq:stationary_examples})
via numerical iteration.
Note that, due to $\tilde{w}[\xi]=\tilde{w}[-\xi]$, we have
$P_\infty(\bk|x )=P_\infty(\bk|-x )$. Hence $f[m]=f[-m]$, and $m=0$ always corresponds to a saddle-point of
$f[m]$. Note also that for $J_{Hb}=0$ (no hydrogen bonding)
considerable further simplification of
(\ref{eq:free_energy_examples},\ref{eq:stationary_examples}) will
be possible, due to the resulting conservation of the symmetry $k_1=k_3$ by the map
(\ref{eq:distribution_def}).

\begin{figure}[t]
\vspace*{12mm}
\setlength{\unitlength}{1.mm}
\begin{picture}(120,60)
\put( 2,  -4){\epsfysize=80\unitlength\epsfbox{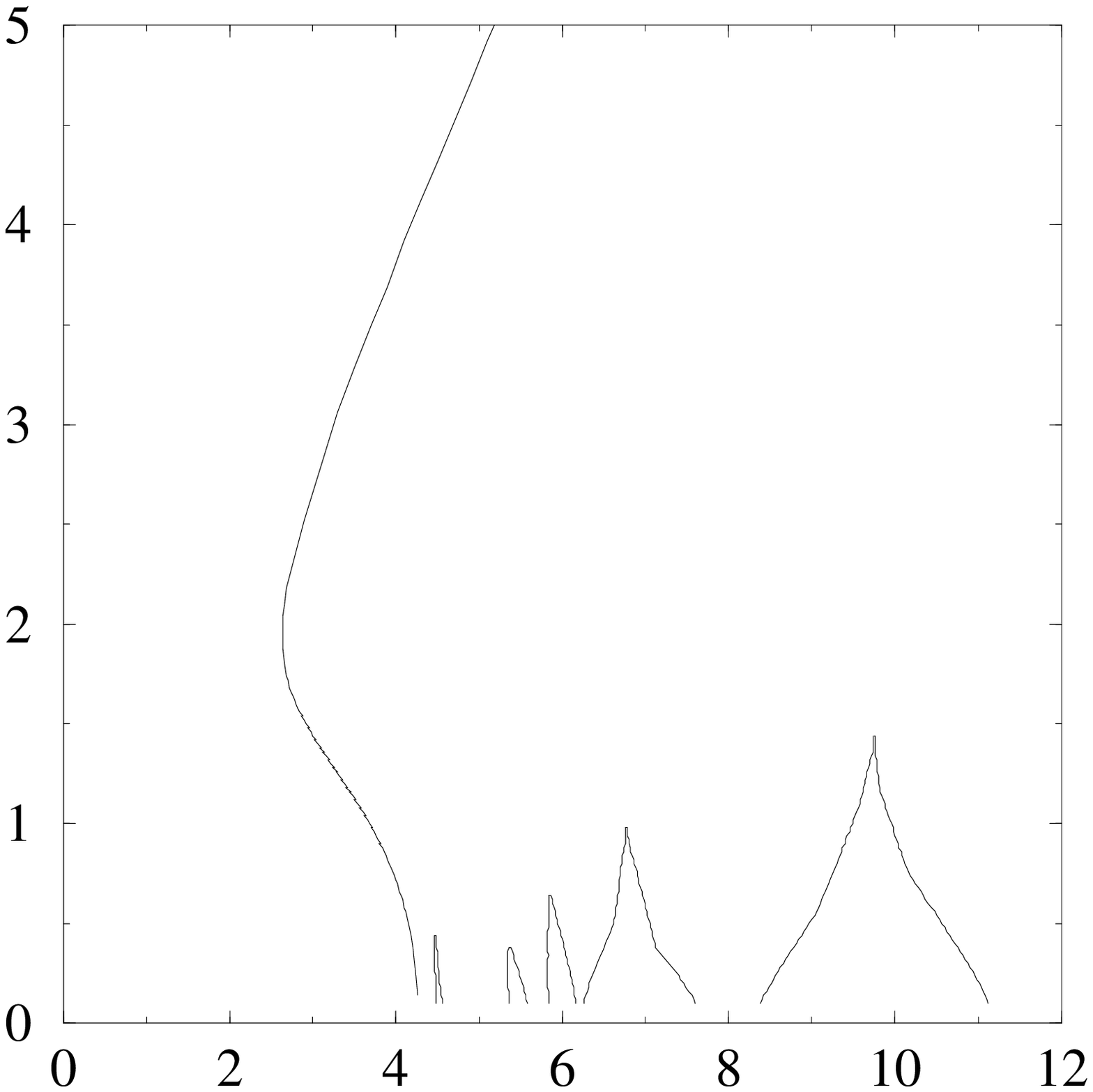}}
\put( 4, 37){$T$}
\put( 53,59){$J_{Hb}=0$}
\put(20,45){\large P}
\put(40,35){\large F}
\put(60.5,11){\large M}
\put( 43,  -1){$J_p$}
\put( 82,  -4){\epsfysize=80\unitlength\epsfbox{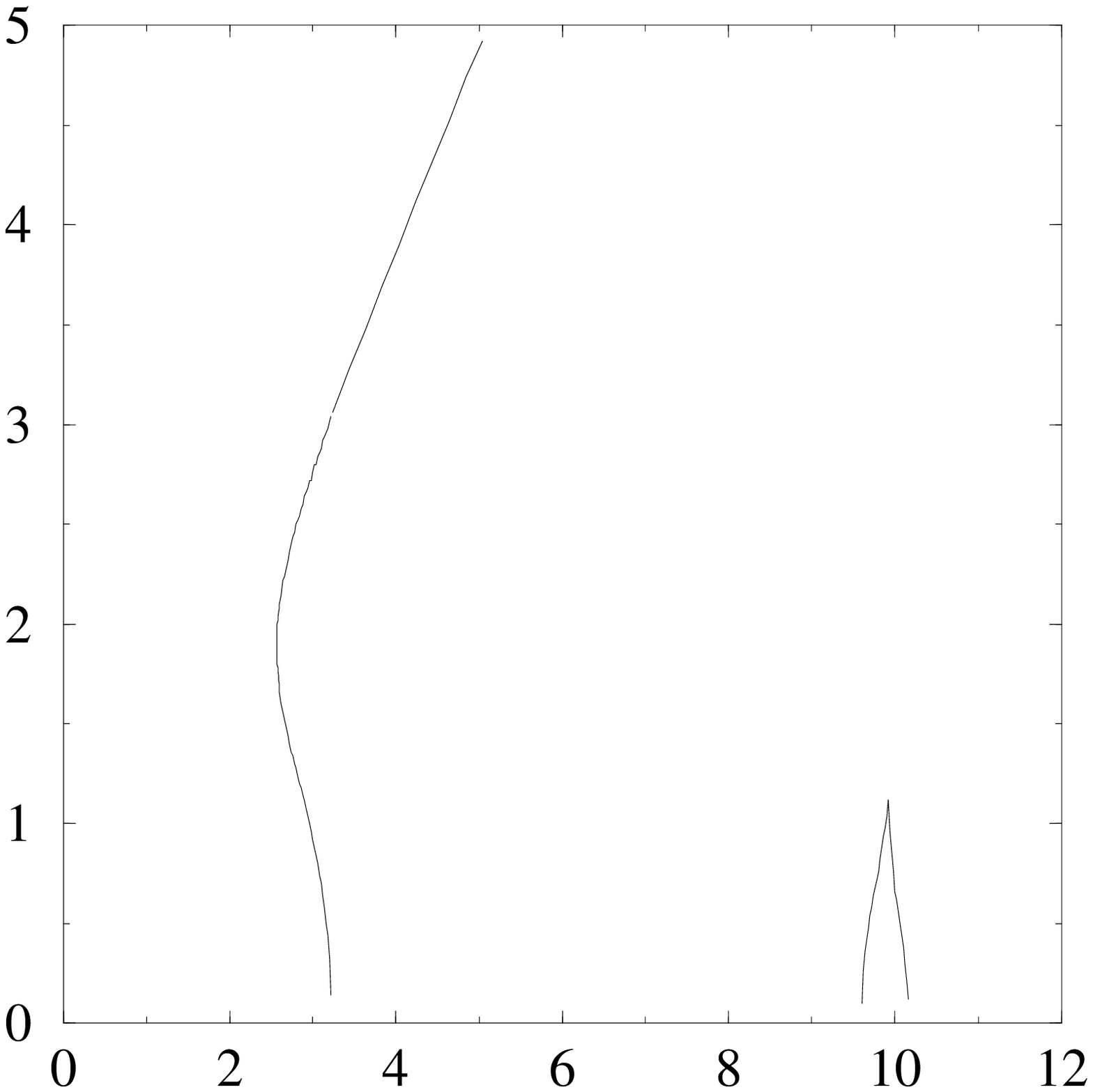}}
\put( 84, 37){$T$}
\put( 130,59){$J_{Hb}=0.5$}
\put(100,45){\large P}
\put(120,35){\large F}
\put(138,11){\large M}
\put(123,  -1){$J_{p}$}
\end{picture}
\vspace*{3mm}
\caption{Phase diagrams cross-sections in the $(J_p,T)$ plane, for $J_s=4$ and $J_{Hb}=0$
(left graph) and $J_{Hb}=2$ (right graph), obtained by numerical
solution of (\ref{eq:stationary_examples}) (which becomes
increasingly complicated as $T\to 0$).
They involve a high-temperature region `P' where $m=0$ is the
only local minimum of $f[m]$, a region `F' where two equivalent $m\neq 0$
solutions (one positive, one negative) minimise $f[m]$. In the
low temperature region a series of `mixed' phases `M' emerge, where multiple states with different degrees
of folding can be simultaneously locally stable (four values for $m$ give local minima).
The P$\to$F transition is second-order. The F$\to$M
transitions are first-order (dynamical) transitions. In the presence of
hydrogen bonds, the M phases are found to be increasingly suppressed (see right picture).}
\label{fig:phase_diagram}
\end{figure}

\begin{figure}[t]
\vspace*{4mm}
\setlength{\unitlength}{1.mm}
\begin{picture}(100,60)
\put( 50,  0){\epsfysize=155\unitlength\epsfbox{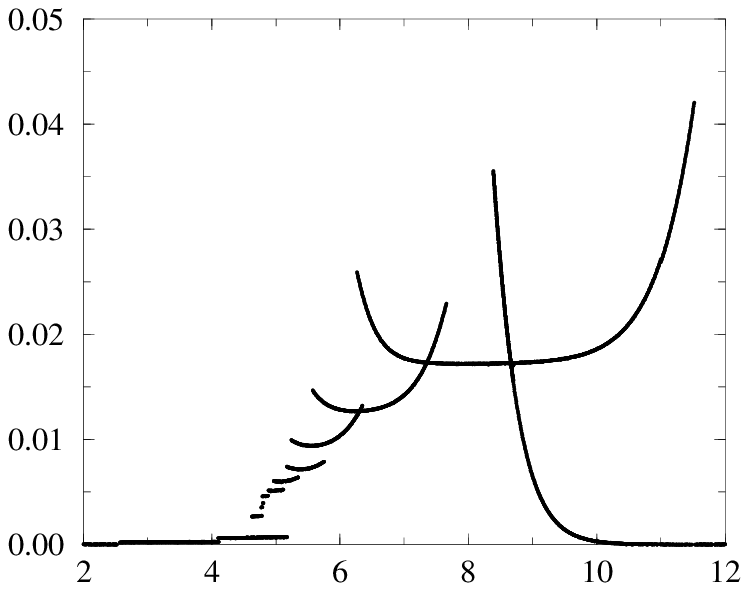}}
\put( 46.5, 32.5){\large $s$}
\put( 90,  -1){$J_p$}
\end{picture}
\vspace*{2mm}
\caption{Entropy per monomer $s=-\partial f/\partial T$ close to zero-temperature (in this graph: $T=0.01$,
$J_s=4$ and $J_{Hb}=0$), as a
function of $J_p$, evaluated numerically via differentiation of the free energy
(\ref{eq:free_energy_final}). It is seen to become non-zero and develop a
hierarchy of sharp peaks at special values of $J_p$, where local frustration is
maximal. `Mixed' phases in the phase diagram emerge at precisely these
locations (see figure \ref{fig:phase_diagram}). }
\label{fig:entropy}
\end{figure}

Examples of the results of our analysis of the surface
(\ref{eq:free_energy_examples}) are shown in figure \ref{fig:phase_diagram},
as phase diagram cross-sections in the $(T,J_p)$ plane, for $\{J_s=4, J_{Hb}=0\}$
(left picture) and $\{J_s=4,J_{Hb}=\frac{1}{2}\}$ (right picture).
They involve
\begin{description}
\item[${\rm (i)}~$]
a high-temperature phase `P', where $m=0$ is the
only local minimum of $f[m]$ and no folding will occur,
\item[${\rm (ii)}~$]
a phase `F' where two equivalent $m\neq 0$
solutions minimise $f[m]$ (one positive, one negative, reflecting the symmetry of the present
model under overall reflection $\phi_i\to \phi_i+\pi$),  the `folded state',
\item[${\rm (iii)}$]
phases `M' where four $m\neq 0$ solutions minimise $f[m]$ locally (two positive, two
negative).
\end{description}
In the M phases, the degree of folding observed will strongly
depend on initial conditions (in spite of the fact that the lowest
value for $f[m]$, and hence the thermodynamic state, corresponds
only to the maximally folded state, where $|m|$ is largest).
See also figure \ref{fig:simulations} below.
The P$\to$F transition is an ordinary second-order transition, whereas the F$\to$M
transitions are first-order (dynamical) transitions. In the presence of
hydrogen bonding, the M phases are found to be increasingly suppressed (see right
picture).

In order to illuminate the
physical mechanism which produces the `mixed' phases, we plot in
figure \ref{fig:entropy} the entropy per monomer $s=-\partial f/\partial T$ close to $T=0$,
for each of the local
minima of $f[m]$.
It is seen to become non-zero, and to develop a
hierarchy of sharp peaks as a function of $J_p$
(c.f. \cite{derridaetal}).
 These peaks correspond to special parameter
values for which frustration effects become dominant, and for which
many energetically equivalent states are possible. The largest
value of the ground state entropy is obtained at the first of these
peaks, for $J_p\approx 11.2$; this corresponds to the location
in the phase diagram where the first of the `mixed' phases appears, see figure
\ref{fig:phase_diagram}.

The qualitative features of diagrams such as those shown in figure
\ref{fig:phase_diagram} can now be
understood as follows. For large values of $\{J_p,T\}$ the short-range forces (steric
forces and hydrogen-bonds) become irrelevant, and
the diagram approaches that of a Mattis model (as it should), with a second-order
transition along the line $T=J_p$. For low temperatures the simple Mattis
state is disrupted  by the steric interactions, which try to enforce
monomer-specific short-range order along the chain; as a result the value needed
for $J_p$ to create $m\neq 0$ states is increased (explaining the
re-entrance observed in figures \ref{fig:phase_diagram}. The
complex phenomenology (reminiscent of random field models) of multiple locally stable
configurations, induced by the
steric interactions, is subsequently found to be damped by the
hydrogen bonds, which act to reduce the complexity of the ground state.

\begin{figure}[t]
\vspace*{3mm}
\setlength{\unitlength}{1.mm}
\begin{picture}(100,60)
\put( 3,  0){\epsfysize=155\unitlength\epsfbox{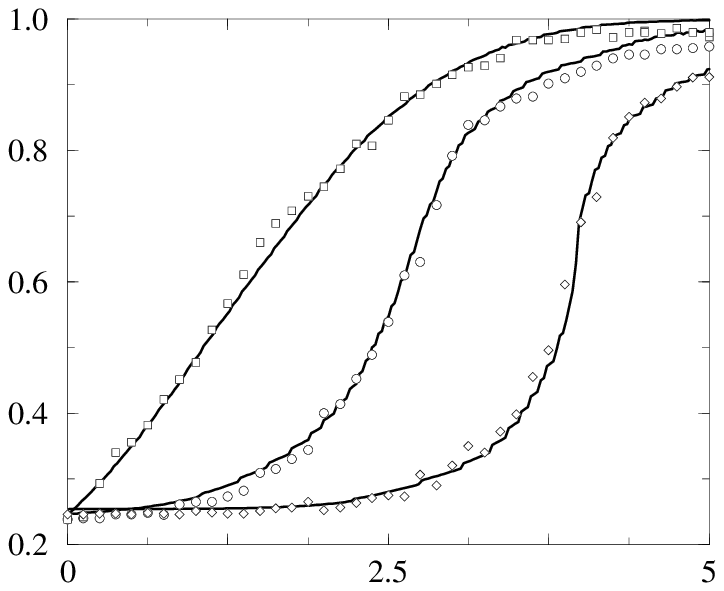}}
\put( 2, 33){$\chi$}
\put( 42,  -1){$J_{Hb}$}
\put( 83,  0){\epsfysize=155\unitlength\epsfbox{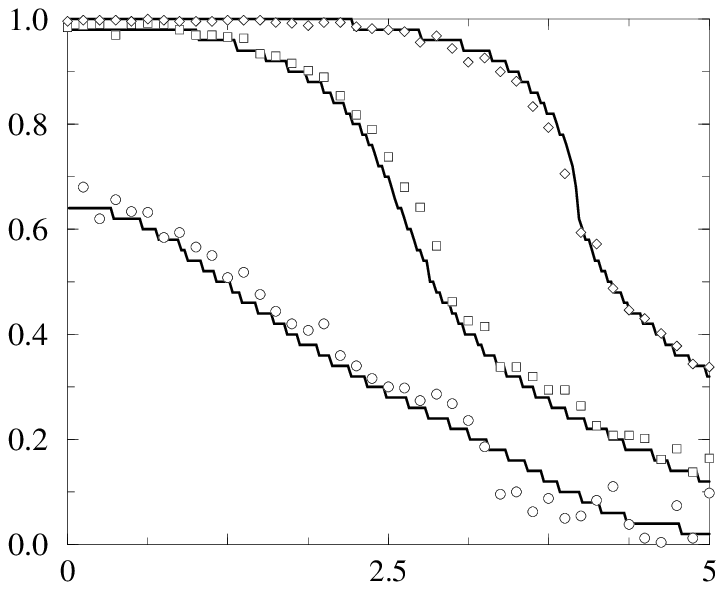}}
\put( 82, 33){$m$}
\put(122,  -1){$J_{Hb}$}
\end{picture}
\vspace*{3mm}
\caption{Equilibrium values of the `chirality'  and `polarity' order parameters $\chi$ and $m$
as functions of the hydrogen-bond strength
$J_{Hb}$. Lines represent theoretical prediction, and markers the simulation
results as measured after $120,000$ iterations per monomer in a system of $N=1000$ monomers.
The values for $J_p$
were chosen as $J_p\in\{4,8,12\}$ (left panel: upper graph to lower graph; right
panel:
lower graph to upper graph).
In all cases  $T=2$ and $J_s=2$. }
\label{fig:observables}
\end{figure}

Next, in figure \ref{fig:observables} we plot the equilibrium values of the `chirality'
(\ref{eq:chirality_ising}) and `polarity' (\ref{eq:polarity_ising}) order
parameters as functions of the hydrogen bond strength $J_{Hb}$, for three different values of $J_p$
(in a region of the phase diagram where there are no mixture
phases, i.e. where apart from overall reflection, the stationary state is unique).
Note that $\chi$ is simply calculated as $\chi=-\frac{1}{2}\partial f/\partial
J_{Hb}$ (which is done numerically).
The two order parameters $\chi$ and $m$ are seen to show an opposite dependence
on $J_{Hb}$
(monotonically increasing vs.\@ decreasing), as they should, since,
$\chi$ measures the degree of helical structure along
the chain, whereas $m$ measures the
probability to find monomers with identical polarity at the same side of the chain.
Due to the competing roles
played by two coupling parameters $\{J_p,J_{Hb}\}$, we see that
`helices' are favoured for large $J_{Hb}$ or small $J_p$ whereas
`folding' in the sense of efficient polarity separation, on the other hand,
is favoured for small $J_{Hb}$ or large
$J_p$. Note that the observed incompatibility of helical structure
with polarity separation is just a reflection of the simple form we chose in this section for
the disorder distribution $\tilde{w}[\eta,\xi]$ (with statistically independent $\eta$ and $\xi$);
the situation would obviously have been different for distributions describing correlated
disorder variables.
In the same figures we also show the results of numerical simulations, for comparison (the markers in the two graphs).
For small $J_{Hb}$ our experiments are seen to be in excellent
agreement with the theory (finite size effects are of the order of
$\mc{O}(N^{-1/2})\approx 0.03$) whereas for large $J_{Hb}$
short-range couplings become increasingly dominant, leading to
domain formation and very slow equilibration times, which make it difficult in practice to probe the
equilibrium regime. In our
experiments we have measured the value of the order parameters
after $120,000$ iterations per spin, which for large $J_{Hb}$ is no longer sufficient.
Note that the theory also predicts the existence of repeated small discontinuous
in both order parameters; these originate from frustration-related short-range
phenomena, as described in e.g. \cite{derridaetal}, which induce discretisation of observable
supports \cite{normand,gyorgyi} and non-analytic integrated distribution
functions (e.g. the Devil's Staircase \cite{bruinsma}).

To verify our results further we have also performed simulation
experiments in the `mixed' phase regions, where our theory predicts that the extent of polarity-driven
folding (i.e. the equilibrium value of $m$) will depend on initial conditions.
In figure \ref{fig:simulations} we
show the value of the `polarity' order parameter $m$, as measured
in numerical simulations of an $N=1000$ chain
after $20,000$ iterations per monomer, as a function of its initial value
$m(t=0)$,
for two different parameter settings (one, to the left, in an M
region of the phase diagram; one, to the right, in an F region of
the phase diagram).
In the insets of these graphs we also plot the corresponding free
energy per monomer $f[m]$ as predicted by our theory, which shows either two $m>0$ locally stable states (left picture)
or one $m>0$ locally stable state (right picture), respectively. In both cases the numerical experiments are
found to verify the existence and the quantitative properties of the expected ergodicity
breaking in the M phase.
We clearly observe that, in phase M, the choice of initial conditions, in particular whether or not
$m(t=0)$ is to the left of the free energy barrier in $f[m]$,
determine the equilibrium value of $m$. We also see that in
the `mixed' phase (left picture) the ergodic
component with the smallest value of $m$ is poorly
equilibrated due to domain formation. This has also been observed for a
similar type of statistical mechanical model in \cite{hebbian}:
in those parameter regions where a multiple number of states can be
locally stable, different ergodic components are found to have
different equilibration time-scales.

\begin{figure}[t]
\vspace*{3mm}
\setlength{\unitlength}{1.mm}
\begin{picture}(100,60)
\put( 3,  0){\epsfysize=155\unitlength\epsfbox{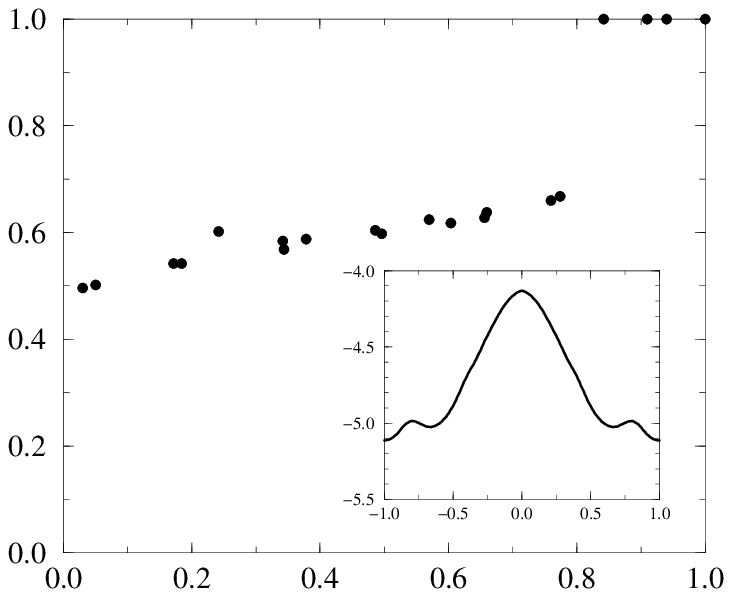}}
\put(  0, 33){$m_{\rm  fin}$}
\put( 41,  -1){$m_{\rm ini}$}
\put(14.23,42){\line(1,0){47.2}}
\put( 85,  0){\epsfysize=155\unitlength\epsfbox{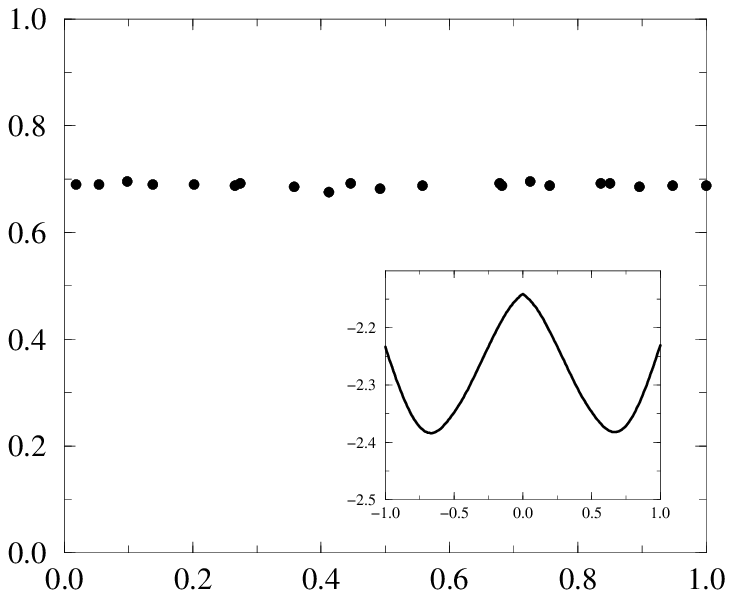}}
\put( 83, 33){$m_{\rm  fin}$}
\put(124,  -1){$m_{\rm ini}$}
\put(96.23,43){\line(1,0){61}}
\end{picture}
\vspace*{3mm}
\caption{Order parameter $m$ as measured in numerical simulations of an $N=1000$
chain after $20,000$ iterations/monomer ($m_{\rm fin}$), versus its initial
value $m_{\rm ini}=m(t=0)$, for system parameters $J_s=4$, $J_p=10$, $J_{Hb}=1$
and $T=0.1$ (left picture, in the M phase) and $J_s=2$, $J_p=4$, $J_{Hb}=1$ and
$T=0.5$ (right picture, in the F phase). In the `folding' phase (right), our
theory predicts the existence of only one $m>0$ ergodic component  (see free
energy per monomer $f[m]$, graph in the inset), at $m\approx 0.67$ (horizontal
line). In the `mixed' phase (left), our theory predicts the existence of two
$m>0$ ergodic components (see free energy per monomer $f[m]$, graph in the
inset), at $m\approx 0.65$ (horizontal line, for $m_{\rm ini}<0.774$) and
$m\approx 1$ (for $m_{\rm ini}>0.774)$. This is confirmed by the numerical
simulations (finite size effects are expected to be of order $\Delta m\approx
N^{-\frac{1}{2}}\approx 0.01$). In the $m\approx 0.65$ state of the mixed phase
(left graph, horizontal line), the system is found not yet to be fully
equilibrated (signaled by a dependence of $m_{\rm fin}$ on $m_{\rm ini}$), due
to domain formation.}
\label{fig:simulations}
\end{figure}

\section{Discussion}

In this paper we have presented an exactly solvable model for secondary structure
formation in random hetero-polymers, consisting of amino-acid monomers which are allowed to
interact in three qualitatively different ways: via (short-range) steric interactions,
via (short-range)  hydrogen-bonding, and via (long-range) polarity-induced forces.
Our strategy was to exploit the one-dimensional nature of the
monomer chain, and to separate questions relating to secondary structure formation
from those relating to tertiary structure formation by taking into
account the effects of the latter only via an effective energy term which
measures the {\em potential} for overall energy reduction by
folding (rather than trying to find the actual state realising this
potential). This allows us to move away from real-space
calculations towards a calculation in $1+\infty$ dimensions, where
the statistical mechanical variables represent the  orientations
of the monomer residues relative to the chain axis.
Solution can now be based on a combination of mean-field and
random transfer-matrix techniques, which
in  one-dimensional models are known to reduce the evaluation of the partition
function to a relatively simple numerical problem.
Due to the presence
of  long-range interactions (via polarity-induced forces), phase transitions
are still possible (and do indeed occur) at finite temperatures.

Our order parameters measure the degree of polarity-induced
collapse of the chain, as well as the degree of helicity along the
chain. The phase diagrams exhibit second-order transitions between
`folded' and `unfolded' states, and, for low temperature and sufficiently strong steric interactions,
 a series of `mixed' phases (separated from the previous ones by discontinuous
transitions) where, in addition to the maximally folded states, specific partially folded states can also
be locally stable.
The latter phases are created at  parameter values for which frustration is maximal, and
where the entropy becomes particularly large.
Although in the present paper we have mostly restricted ourselves
(for simplicity) to chains with just a small number of
possible orientations per monomer, it is not fundamentally more difficult to solve
the model for larger degrees of orientational freedom (although
certain adaptations are needed before the continuum limit can be
taken, such as a re-scaling of the effective long range coupling $J_p$ and/or of the
number of relative monomer orientations where polarity interactions occur).
We have only evaluated our theory for the simplest choice of
disorder statistics (the statistical properties of the monomers, and their physical
properties such as polarity and steric constraints).
Here the emerging picture is already quite satisfactory, in that
explicit analytical results can be obtained, and that the predicted physical
behaviour of the monomer chain (confirmed qualitatively and quantitatively
by numerical simulations)
makes perfect sense in the context of proteins: the polarity
forces drive the transition to a collapsed state, the steric
forces introduce monomer specificity, and the hydrogen bonds
stabilise the conformation by damping the frustration-induced
multiplicity of states.

There is still much scope
for increasing the biological realism and relevance of our model without affecting
its analytical solvability,
at different levels. Firstly, without changing the model or its techniques
for solution, one can easily consider more realistic choices for the
monomer statistics, such us non-binary polarity variables, or for the orientational freedom of the
monomers (for instance, the hydrogen-bond term may be modified to
favour helix-type formations at the biologically observed ratio of 3.6 monomers
per turn). Secondly, at a next level of sophistication one could construct a more realistic
form for the polarity induced energy contribution (breaking the
present hydrophobic-hydrophilic symmetry, and based upon biological data), or more realistic representations
of the degrees of freedom of the individual peptide units and residues (i.e. three
angles per monomer, rather than one), or the action of `chaperones' (via external
fields).
Solution of such models would not be essentially more difficult
than that of the examples worked out here; the main problem would rather be to extract the canonical
definitions of the ingredients to be incorporated into the model from the
available biological data.
In contrast, qualitatively different and more difficult types of modification and extension would be to
consider non-random hetero-polymers, where the monomer properties
and statistics are chosen such as to mimic real proteins, or to
try to analyse the interplay between secondary and tertiary
structure formation. Here new techniques for solution will have to come in.

The main problem in the statistical mechanical study of folding
proteins appears to be the construction of models where  an acceptable and productive  balance can be found
between analytical solvability and
biological realism. We believe that our present model might point to a
new direction where this might be achieved.

\section*{References}


\end{document}